\pdfoutput=1  % serve per imporgli di andare in pdflatex, utile quando ci sono figure grosse in .eps, e arXiv chiede di convertirle in .pdf per farlo %compilare in pdflatex
\def\cO{\cos\Omega}
\def\sO{\sin\Omega}
\def\cI{\cos I}
\def\sI{\sin I}
\def\nk{n_\mathrm{Kep}}
\def\acap{\\ \nonumber \\}

\def\kl{\boldsymbol{\hat{J}}\boldsymbol{\cdot}\boldsymbol{\hat{l}}}
\def\km{\boldsymbol{\hat{J}}\boldsymbol{\cdot}\boldsymbol{\hat{m}}}
\def\kh{\boldsymbol{\hat{J}}\boldsymbol{\cdot}\boldsymbol{\hat{h}}}
\def\Pb{P_\mathrm{Kep}}
\def\rfr#1{Equation\,(\ref{#1})}
\def\rfrs#1#2{Equations\,(\ref{#1})--(\ref{#2})}
\def\Rfr#1{Equation\,(\ref{#1})}

\def\derp#1#2{\rp{\partial{#1}}{\partial{#2}}}
\def\dert#1#2{\frac{{{\textrm{d}}}{#1}}{{{\textrm{d}}}{#2}}}
\def\virg#1{``#1"}
\def\eqi{\begin{equation}}
\def\eqf{\end{equation}}
\def\rp#1#2{\frac{#1}{#2}}
\def\lb#1{\label{#1}}
\def\bds#1{\mathbf{#1}}
\def\ton#1{\left(#1\right)}
\def\qua#1{\left[#1\right]}
\def\grf#1{\left\{#1\right\}}

\RequirePackage[2020-02-02]{latexrelease}
\documentclass{aastex631}
\usepackage{morefloats}
\usepackage[title]{appendix}
\usepackage{textcomp}
\usepackage{booktabs}
\usepackage{multirow}
\usepackage{rotating,tabularx}
\usepackage{float}
\usepackage{enumerate}
\usepackage{rotating}
\usepackage[polutonikogreek,english]{babel}
\usepackage{amsmath,starfont,textgreek,w-greek}
\usepackage[flushleft]{threeparttable}
\usepackage{amsthm}
\usepackage{amscd}
\usepackage[mathlines]{lineno}
\usepackage{amssymb,dsfont}
\usepackage{graphicx,epsfig}
\usepackage{txfonts}
\usepackage{xr-hyper}
\usepackage{hyperref}

\usepackage[nointegrals]{wasysym}
\usepackage[caption=false]{subfig}

\usepackage{tikz,tikz-3dplot}
\usetikzlibrary{decorations.markings, calc, fadings,
decorations.pathreplacing, patterns, decorations.pathmorphing, positioning}

\newcommand{\AxisRotator}[1][rotate=0] {%
  \tikz[decoration={
    markings,
    mark=at position 1 with ->}]\draw[x = .5em, y = 2.75em, line width = .2ex,#1,postaction=decorate] (0,0)  arc (-150:150:.45 and .35) -- ++(-95:2pt);%
  }

\newcommand{\tdseteulerxyz}{
\renewcommand{\tdplotcalctransformrotmain}{%
\tdplotsinandcos{\sinalpha}{\cosalpha}{\tdplotalpha}
\tdplotsinandcos{\sinbeta}{\cosbeta}{\tdplotbeta}
\tdplotsinandcos{\singamma}{\cosgamma}{\tdplotgamma}
\tdplotmult{\sasb}{\sinalpha}{\sinbeta}
\tdplotmult{\sasg}{\sinalpha}{\singamma}
\tdplotmult{\sasbsg}{\sasb}{\singamma}
\tdplotmult{\sacb}{\sinalpha}{\cosbeta}
\tdplotmult{\sacg}{\sinalpha}{\cosgamma}
\tdplotmult{\sasbcg}{\sasb}{\cosgamma}
\tdplotmult{\casb}{\cosalpha}{\sinbeta}
\tdplotmult{\cacb}{\cosalpha}{\cosbeta}
\tdplotmult{\cacg}{\cosalpha}{\cosgamma}
\tdplotmult{\casg}{\cosalpha}{\singamma}
\tdplotmult{\cbsg}{\cosbeta}{\singamma}
\tdplotmult{\cbcg}{\cosbeta}{\cosgamma}
\tdplotmult{\casbsg}{\casb}{\singamma}
\tdplotmult{\casbcg}{\casb}{\cosgamma}
\pgfmathsetmacro{\raaeul}{\cacb}
\pgfmathsetmacro{\rabeul}{\casbsg - \sacg}
\pgfmathsetmacro{\raceul}{\sasg + \casbcg}
\pgfmathsetmacro{\rbaeul}{\sacb}
\pgfmathsetmacro{\rbbeul}{\sasbsg + \cacg}
\pgfmathsetmacro{\rbceul}{\sasbcg - \casg}
\pgfmathsetmacro{\rcaeul}{-\sinbeta}
\pgfmathsetmacro{\rcbeul}{\cbsg}
\pgfmathsetmacro{\rcceul}{\cbcg}
}
}

\definecolor{amber}{rgb}{1.0, 0.75, 0.0}
\definecolor{ao(english)}{rgb}{0.0, 0.5, 0.0}
\definecolor{antiquebrass}{rgb}{0.8, 0.58, 0.46}
\definecolor{bole}{rgb}{0.47, 0.27, 0.23}
\definecolor{uclagold}{rgb}{1.0, 0.7, 0.0}
\allowdisplaybreaks

\makeatletter
 \DeclareRobustCommand\ref{%
    \@ifstar\@refstar\T@ref
  }%
  \DeclareRobustCommand\pageref{%
    \@ifstar\@pagerefstar\T@pageref
  }%
 \makeatother

\begin{document}

\title{Revisiting the gravitomagnetic clock effect}

\shortauthors{L. Iorio}

\author[0000-0003-4949-2694]{Lorenzo Iorio}
\affiliation{Ministero dell' Istruzione e del Merito. Viale Unit\`{a} di Italia 68, I-70125, Bari (BA),
Italy}

\email{lorenzo.iorio@libero.it}

\begin{abstract}
To the first post-Newtonian order, if two test particles revolve in opposite directions about a massive, spinning body along two circular and equatorial orbits with the same radius, they take different times to return to the reference direction relative to which their motion is measured: it is the so-called gravitomagnetic clock effect. The satellite moving in the same sense of the rotation of the primary is slower, and experiences a retardation with respect to the case when the latter does not spin, while the one circling  in the opposite sense of the rotation of the source is faster, and its orbital period is shorter than it would be in the static case. The resulting time difference due to the stationary gravitomagnetic field of the central spinning body is proportional to the angular momentum per unit mass of the latter through a numerical factor which so far has been found to be  $4\pi$. A numerical integration of the equations of motion of a fictitious test particle moving along a circular path lying in the equatorial plane of a hypothetical rotating object by including the gravitomagnetic acceleration to the first post-Newtonian order shows that, actually, the gravitomagnetic corrections to the orbital periods  are larger by a factor of $4$ in both the prograde and retrograde cases. Such an outcome, which makes the proportionality coefficient of the gravitomagnetic difference in the orbital periods of the two counter-revolving orbiters equal to $16\pi$, confirms an analytical calculation recently published in the literature by the present author. \textcolor{black}{It is an important result in view of the astrophysical implications of the gravitomagnetic clock effect around Kerr black holes.}
\end{abstract}

%{
%\textrm{Unified Astronomy Thesaurus concepts}:\,Exoplanets\,(498); General relativity\,(641)
%}
%\centerline
%{PACS: 04.20.-q; 04.20.Cv; 04.80.-y; 04.80.Cc; 91.10.Sp}

\section{Introduction}
Within the weak-field and slow-motion approximation of the General Theory of Relativity (GTR), by gravitomagnetic clock effect it is usually  meant the difference $\Delta T_\mathrm{gvm}$ between the orbital periods of two counter-revolving test particles\textcolor{black}{, i.e. non-spinning,} moving along circular orbits of identical radius $r_0$ in the equatorial plane of a rotating body of mass $M$ and angular momentum $\boldsymbol{J}$ \citep{MitPG,Vladi87,1993PhLA..181..353C,Mash97,You98,Mashetal99,Tart00,MashSan00,Mashetal01,2014PhRvD..90d4059H,2023arXiv231112018S}. It turns out that $\Delta T_\mathrm{gvm}$ is proportional to $J/\ton{M\,c^2}$, where $c$ is the speed of light in vacuum, through a numerical factor that has been calculated in the literature to be equal to $4\pi$. Such an intriguing relativistic feature of motion was the subject of several papers investigating its possible detection \textcolor{black}{as well}; see \citet{Gron97,Mashetal99,2000AdSpR..25.1255L,2000CQGra..17..783T,2000CQGra..17.2381T,2001IJMPD..10..465I,2001CQGra..18.4303I,2005CQGra..22..119I,2006AnP...518..868L}.
\textcolor{black}{It has also relevant consequences in astrophysical contexts such as Kerr black hole spacetimes \citep{1995CQGra..12.1119D,1999CQGra..16.1853B,2002Napoli,2004PhLA..327...95F,2005CQGra..22.4729B}. For other versions of the gravitomagnetic clock effect involving spinning orbiters in the Kerr spacetime, see, e.g., \citet{2004CQGra..21.5441B,2006PhRvD..74l4006M}.}

The standard approach in deriving the aforementioned form of the gravitomagnetic clock effect is to calculate the time interval $T_\mathrm{gvm}$ required to a test particle to come back to some fixed reference direction in the orbital plane from which it began its motion, assumed circular throughout the overall variation of the azimuthal angle $\upvarphi$ reckoned from such a line and spanning an interval of $2\pi$, when the general relativistic gravitomagnetic acceleration is added to the Newtonian inverse-square one. The unit vector $\boldsymbol{\hat{J}}$ of the primary's angular momentum is assumed to be known, so that one can align the reference $z$ axis with it, and the reference $\grf{x,\,y}$ plane coincides with the equatorial one of the source. \textcolor{black}{It is assumed the point of view of a distant observer fixed with respect to distant stars who uses the coordinate time $t$ as own proper time; the difference of the proper times $\uptau$ of the counter-orbiting test particles is identical to $T_\mathrm{gvm}$ up to corrections of order $\mathcal{O}\ton{c^{-4}}$. Indeed, it is  $\mathrm{d}\uptau =\sqrt{g_{00}}\,\mathrm{d}t\simeq \sqrt{1+h_{00}}\,\mathrm{d}t\simeq \ton{1 + h_{00}/2}\,\mathrm{d}t$, where $g_{00}$ is the $\virg{00}$ component of the spacetime metric tensor, and $h_{00}$ is a small correction of post-Newtonian (pN) order.}

In this paper, we will numerically check such an established result by integrating the equations of motion of a fictitious test particle orbiting a putative spinning body by removing the foregoing restrictions about $\boldsymbol{\hat{J}}$. Stated differently, a circular and equatorial orbit will be considered, but its orientation in the reference frame adopted will be arbitrary. It turns out that, in such a scenario, the line of the nodes, among other things, remains fixed; then, it will be naturally assumed as reference polar axis from which the azimuthal angle $\upvarphi$ is counted in the orbital plane. Furthermore, the argument of latitude $u$, which reckons the instantaneous position of the test particle along its orbit just from the unit vector $\boldsymbol{\hat{l}}$ of the line of the nodes, will play the role of  $\upvarphi$. Thus, the time interval\footnote{In the general case, it is known as draconitic period. See \citet{Capde05} or Footnote 1 of \citet{2016MNRAS.460.2445I} for the origin of such a name.} between two consecutive crossings of the fixed line of the nodes  by the test particle will be inspected by looking at the time needed to the cosine of the angle between its position vector $\boldsymbol{r}$ and $\boldsymbol{\hat{l}}$ to assume again its initial value. The numerical integration will be performed for both the senses of motion of the test particle. As a result, a discrepancy larger than the established form of the gravitomagnetic clock effect by a multiplicative factor of\footnote{See Section 4 of \citet{2006AnP...518..868L} for an alternative clock effect in agreement with the present treatment, despite less general.} $4$ will be found, in agreement with an analytical calculation recently appeared in the literature \citep{2016MNRAS.460.2445I}.

The paper is organized as follows. In Section\,\ref{sec:2}, the gravitomagnetic acceleration experienced by a satellite orbiting a rotating primary and some of its main orbital effects are reviewed. The gravitomagnetic features of circular and equatorial orbits arbitrarily oriented in space are outlined in Section\,\ref{sec:3}. Section\,\ref{sec:4} is devoted to the numerical integrations of the equations of motion in the scenario of Section\,\ref{sec:3}. Section\,\ref{sec:5} deals with a tentative explanation of the resulting discrepancy with respect to the standard case. The impact of the gravitomagnetic field on the interval between two consecutive passages of the test particle at the periapsis, known as anomalistic period, is numerically investigated in Section\,\ref{sec:6}. The possibility of using the existing Earth's artificial satellites LAGEOS and LARES 2 to detect a certain form of gravitomagnetic clock effect in view of their peculiar orbital configuration is investigated in Section\,\ref{sec:7}. Section\,\ref{sec:8} summarizes the obtained results and offers concluding remarks. Appendix\,\ref{appenA} includes a list of the symbols and definitions used in the paper.
\section{The Lense-Thirring acceleration and some of its orbital consequences}\lb{sec:2}
To the first post-Newtonian (1pN) order, the mass-energy currents of an isolated, spinning body perturb the orbital motion of a test particle which experiences the gravitomagnetic Lense-Thirring (LT) acceleration \citep{Sof89,iers10,2014grav.book.....P,SoffelHan19}
\eqi
{\bds A}^\mathrm{LT} = \rp{2\,G\,J}{c^2\,r^3}\,\qua{3\,\ton{\boldsymbol{\hat{J}}\boldsymbol{\cdot}\boldsymbol{\hat{r}}}\,\boldsymbol{\hat{r}}\boldsymbol{\times}\boldsymbol{v} + \boldsymbol{v}\boldsymbol{\times}\boldsymbol{\hat{J}} }\lb{ALT}
\eqf
in addition to the Newtonian one.

The $R-T-N$ components of \rfr{ALT} turn out to be
\begin{align}
\mathrm{A}_R^\mathrm{LT} \lb{ARLT} & = \rp{2\,\nk\,G\,J\,\ton{1 + e\,\cos f}^4\,\ton{\kh}}{c^2\,a^2\,\ton{1 - e^2}^{7/2}}, \acap
\mathrm{A}_T^\mathrm{LT} \lb{ATLT} & = -\rp{2\,e\,\nk\,G\,J\,\ton{1 + e\,\cos f}^3\sin f\,\ton{\kh}}{c^2\,a^2\,\ton{1 - e^2}^{7/2}}, \acap
\mathrm{A}_N^\mathrm{LT} \lb{ANLT} & = -\rp{2\,\nk\,G\,J\,\ton{1 + e\,\cos f}^3}{c^2\,a^2\,\ton{1 - e^2}^{7/2}}\,\grf{
\qua{e \,\cos\omega - \ton{2 + 3\,e\,\cos f}\,\cos u}\,\ton{\kl} -
 \rp{1}{2}\,\qua{e\,\sin\omega + 4\,\sin u + 3\,e\,\sin\ton{2f + \omega}}\,\ton{\km}
}.
\end{align}

\Rfr{ALT}, through \rfrs{ARLT}{ANLT}, induces the LT effect  \citep{1918PhyZ...19..156L,1984GReGr..16..711M} consisting of the following secular precessions of some Keplerian orbital elements \citep{1975PhRvD..12..329B,1988NCimB.101..127D,1992PhRvD..45.1840D,1999ApJ...514..388W,2007CQGra..24.1775K,2008ApJ...674L..25W,2017EPJC...77..439I}
\begin{align}
\dert I t \lb{Irate}& = \rp{2\,G\,J\,\ton{\kl}}{c^2\,a^3\,\ton{1-e^2}^{3/2}}, \\ \nonumber\\
\dert \Omega t \lb{Orate}& = \rp{2\,G\,J\,\csc I\,\ton{\km}}{c^2\,a^3\,\ton{1-e^2}^{3/2}}, \\ \nonumber\\
\dert \omega t \lb{orate}& = -\rp{2\,G\,J\,\qua{2\,\ton{\kh} + \cot I\,\ton{\km}}}{c^2\,a^3\,\ton{1-e^2}^{3/2}}.
\end{align}
They are currently under measurement with some Earth's geodetic satellites \citep{2019JGeod..93.2181P} tracked with the Satellite Laser Ranging (SLR) technique \citep{SLR11}; see, e.g., \citep{2011Ap&SS.331..351I,2013CEJPh..11..531R,2020Univ....6..139L}, and references therein.

A further gravitomagnetic effect, i.e. the spin precessions of orbiting gyroscopes \citep{Pugh59,Schiff60}, was recently measured in the Earth's field with the  Gravity Probe B (GP-B) mission to a $19\%$ accuracy level \citep{2011PhRvL.106v1101E,2015CQGra..32v4001E,2011PhyOJ...4...43W}; the originally expected uncertainty was $\simeq 1\%$ \citep{Varenna74,2001LNP...562...52E}.
\section{Gravitomagnetic effects for a circular and equatorial orbit}\lb{sec:3}
Let us consider an orbital configuration such that the satellite's angular momentum $\boldsymbol{h}$ is (anti-)parallel to the body's spin axis, irrespectively of the orientation of the latter one in the reference frame adopted,  i.e. such that
%\footnote{It turns out that \rfrs{kh}{klm} are fulfilled for $I = 90^\circ - \delta_J,\,\Omega = 90^\circ + \alpha_J$.}
\begin{align}
\boldsymbol{\hat{J}}\boldsymbol{\cdot}\boldsymbol{\hat{h}} \lb{kh}& =\pm 1,\\ \nonumber \\
\boldsymbol{\hat{J}}\boldsymbol{\cdot}\boldsymbol{\hat{l}}\lb{klm}& =\boldsymbol{\hat{J}}\boldsymbol{\cdot}\boldsymbol{\hat{m}} = 0;
\end{align}
see Figure\,\ref{fig:pro} and Figure\,\ref{fig:retro}.
%
%%%%%%%%%%%%%%%%%%%%%%%%%%%%%%%%%%%%%%%%%%%%%%%%%%%%%%%%%%%%%%%%%%%%%%%%%%%%%%%%%%%%%%%%%%%%%%%%%%%%%%%%%%
%%%%%%%%%%%%%%%%%%%%%%%%%%%%%%%%%%%%%%%%%%%%%%%%%%%%%%%%%%%%%%%%%%%%%%%%%%%%%%%%%%%%%%%%%%%%%%%%%%%%%%%%%%
\tdplotsetmaincoords{70}{110}
\tdseteulerxyz
\begin{figure}
\centering
\begin{tikzpicture}[tdplot_main_coords,scale=7]
  \pgfmathsetmacro{\RA}{150} % RA of the spin axis
  \pgfmathsetmacro{\DEC}{75} % DEC of the spin axis
  \pgfmathsetmacro{\a}{1} % semimajor axis
  \pgfmathsetmacro{\e}{0.0} % eccentricity
  \pgfmathsetmacro{\i}{30} % inclination [deg]
  \pgfmathsetmacro{\O}{45} % right ascension of ascending node [deg]
  \pgfmathsetmacro{\o}{50} % argument of pericenter [deg]
  \pgfmathsetmacro{\fo}{60} % true anomaly [deg]
  \pgfmathsetmacro{\fpa}{0} % true anomaly at periapsis [deg]
  \pgfmathsetmacro{\faa}{180} % true anomaly at apoapsis [deg]
  \pgfmathsetmacro{\fan}{360-\o} % true anomaly at ascending node [deg]
  \pgfmathsetmacro{\fdn}{180-\o} % true anomaly at descending node [deg]
  \pgfmathsetmacro{\fm}{90-\o} % true anomaly at m [deg]
  \pgfmathsetmacro{\apl}{1} % semimajor axis of the fundamental plane
  \pgfmathsetmacro{\epl}{0} % eccentricity of the fundamental plane
  \pgfmathsetmacro{\ipl}{0} % inclination of the fundamental plane [deg]
  \pgfmathsetmacro{\Opl}{0} % right ascension of ascending node of the fundamental plane [deg]
  \pgfmathsetmacro{\opl}{0} % argument of pericenter of the fundamental plane [deg]
  \pgfmathsetmacro{\scl}{0.75} % scaling parameter for the dotted part of line of apsides below the fundamental plane
  \pgfmathsetmacro{\s}{1.1} % scaling parameter for legends
  \pgfmathsetmacro{\ofs}{0.03} % offset parameter for legends
  \pgfmathsetmacro{\u}{1} % offset parameter for the unit vectors
   \pgfmathsetmacro{\uJ}{0.35} % offset parameter for the unit vector J
   \pgfmathsetmacro{\usJ}{0.27} % offset parameter for the unit vector J
  \pgfmathsetmacro{\k}{0.3} % offset parameter for the RTN unit vectors
   \pgfmathsetmacro{\um}{0.8} % offset parameter for the m unit vector

%---------------------Position of the origin of the coordinates --------------------------------

\coordinate (O) at (0,0,0);

%-------------------Position of the test particle ------------------------------------

\coordinate (P) at (
{(\a*(1-\e^2)/(1+\e*cos(\fo)))*(cos(\O)*cos(\o+\fo)-cos(\i)*sin(\O)*sin(\o+\fo))},
{(\a*(1-\e^2)/(1+\e*cos(\fo)))*(sin(\O)*cos(\o+\fo)+cos(\i)*cos(\O)*sin(\o+\fo))},
{(\a*(1-\e^2)/(1+\e*cos(\fo)))*(sin(\i)*sin(\o+\fo))}
);

%------------------------- Position of the legend of the test particle ----------------------

\coordinate (rP) at (
{\s*(\a*(1-\e^2)/(1+\e*cos(\fo)))*(cos(\O)*cos(\o+\fo)-cos(\i)*sin(\O)*sin(\o+\fo))},
{\s*(\a*(1-\e^2)/(1+\e*cos(\fo)))*(sin(\O)*cos(\o+\fo)+cos(\i)*cos(\O)*sin(\o+\fo))-0.25},
{\s*(\a*(1-\e^2)/(1+\e*cos(\fo)))*(sin(\i)*sin(\o+\fo))-0.2}
);

%------------------------------ Position of the ascending node ----------------------------------------

\coordinate (AN) at (
{(\a*(1-\e^2)/(1+\e*cos(\fan)))*(cos(\O)*cos(\o+\fan)-cos(\i)*sin(\O)*sin(\o+\fan))},
{(\a*(1-\e^2)/(1+\e*cos(\fan)))*(sin(\O)*cos(\o+\fan)+cos(\i)*cos(\O)*sin(\o+\fan))},
{(\a*(1-\e^2)/(1+\e*cos(\fan)))*(sin(\i)*sin(\o+\fan))}
);

%-------------------------------- Position of the legend of the ascending node -----------------------

\coordinate (rAN) at (
{\s*(\a*(1-\e^2)/(1+\e*cos(\fan)))*(cos(\O)*cos(\o+\fan)-cos(\i)*sin(\O)*sin(\o+\fan))},
{\s*(\a*(1-\e^2)/(1+\e*cos(\fan)))*(sin(\O)*cos(\o+\fan)+cos(\i)*cos(\O)*sin(\o+\fan))},
{\s*(\a*(1-\e^2)/(1+\e*cos(\fan)))*(sin(\i)*sin(\o+\fan))}
);

%-------------------------------- Position of the end of the line of nodes in the fundamental plane -----------------------

\coordinate (lAN) at (
{\apl*(\a*(1-\e^2)/(1+\e*cos(\fan)))*(cos(\O)*cos(\o+\fan)-cos(\i)*sin(\O)*sin(\o+\fan))},
{\apl*(\a*(1-\e^2)/(1+\e*cos(\fan)))*(sin(\O)*cos(\o+\fan)+cos(\i)*cos(\O)*sin(\o+\fan))},
{\apl*(\a*(1-\e^2)/(1+\e*cos(\fan)))*(sin(\i)*sin(\o+\fan))}
);

%------------------------------Position of the legend of the unit vector l ---------------------------

\coordinate (rl) at (
{\s*(\a*(1-\e^2)/(1+\e*cos(\fan)))*(cos(\O)*cos(\o+\fan)-cos(\i)*sin(\O)*sin(\o+\fan))},
{\s*(\a*(1-\e^2)/(1+\e*cos(\fan)))*(sin(\O)*cos(\o+\fan)+cos(\i)*cos(\O)*sin(\o+\fan))},
{\s*(\a*(1-\e^2)/(1+\e*cos(\fan)))*(sin(\i)*sin(\o+\fan))+\ofs}
);

%----------------------- Position of the symbol of the ascending node -------------------------

\coordinate (rasc) at (
{\s*(\a*(1-\e^2)/(1+\e*cos(\fan)))*(cos(\O)*cos(\o+\fan)-cos(\i)*sin(\O)*sin(\o+\fan))},
{\s*(\a*(1-\e^2)/(1+\e*cos(\fan)))*(sin(\O)*cos(\o+\fan)+cos(\i)*cos(\O)*sin(\o+\fan))},
{\s*(\a*(1-\e^2)/(1+\e*cos(\fan)))*(sin(\i)*sin(\o+\fan))-\ofs}
);

%------------------------- Position of the legend of the line of nodes --------------------------------

\coordinate (lon) at (
{(\a*(1-\e^2)/(1+\e*cos(\fan)))*(cos(\O)*cos(\o+\fan)-cos(\i)*sin(\O)*sin(\o+\fan))},
{1.5*\s*(\a*(1-\e^2)/(1+\e*cos(\fan)))*(sin(\O)*cos(\o+\fan)+cos(\i)*cos(\O)*sin(\o+\fan))},
{(\a*(1-\e^2)/(1+\e*cos(\fan)))*(sin(\i)*sin(\o+\fan))-\ofs}
);

%-------------------------- Position of the descending node ---------------------------------

\coordinate (DN) at (
{(\a*(1-\e^2)/(1+\e*cos(\fdn)))*(cos(\O)*cos(\o+\fdn)-cos(\i)*sin(\O)*sin(\o+\fdn))},
{(\a*(1-\e^2)/(1+\e*cos(\fdn)))*(sin(\O)*cos(\o+\fdn)+cos(\i)*cos(\O)*sin(\o+\fdn))},
{(\a*(1-\e^2)/(1+\e*cos(\fdn)))*(sin(\i)*sin(\o+\fdn))}
);

%------------------------ Position of the legend of the descending node--------------------------------

\coordinate (rDN) at (
{\s*(\a*(1-\e^2)/(1+\e*cos(\fdn)))*(cos(\O)*cos(\o+\fdn)-cos(\i)*sin(\O)*sin(\o+\fdn))},
{\s*(\a*(1-\e^2)/(1+\e*cos(\fdn)))*(sin(\O)*cos(\o+\fdn)+cos(\i)*cos(\O)*sin(\o+\fdn))},
{\s*(\a*(1-\e^2)/(1+\e*cos(\fdn)))*(sin(\i)*sin(\o+\fdn))}
);

%---------------------- Position of the symbol of the descending node------------------------

\coordinate (rdesc) at (
{\s*(\a*(1-\e^2)/(1+\e*cos(\fdn)))*(cos(\O)*cos(\o+\fdn)-cos(\i)*sin(\O)*sin(\o+\fdn))},
{\s*(\a*(1-\e^2)/(1+\e*cos(\fdn)))*(sin(\O)*cos(\o+\fdn)+cos(\i)*cos(\O)*sin(\o+\fdn))},
{\s*(\a*(1-\e^2)/(1+\e*cos(\fdn)))*(sin(\i)*sin(\o+\fdn))+\ofs}
);

%---------------------------------- Position of the unit vector h ---------------------------------

\coordinate (h) at (
{\u*sin(\i)*sin(\O)},
{-\u*sin(\i)*cos(\O)},
{\u*cos(\i)}
);

%---------------------- Position of the legend of the unit vector h ------------------------

\coordinate (rh) at (
{\s*\u*sin(\i)*sin(\O)},
{-\s*\u*sin(\i)*cos(\O)},
{\s*\u*cos(\i)}
);

%-----------------------------------Position of the unit vector m -------------------------

\coordinate (m) at (
{\um*(\a*(1-\e^2)/(1+\e*cos(\fm)))*(cos(\O)*cos(\o+\fm)-cos(\i)*sin(\O)*sin(\o+\fm))},
{\um*(\a*(1-\e^2)/(1+\e*cos(\fm)))*(sin(\O)*cos(\o+\fm)+cos(\i)*cos(\O)*sin(\o+\fm))},
{\um*(\a*(1-\e^2)/(1+\e*cos(\fm)))*(sin(\i)*sin(\o+\fm))}
);

%------------------------------ Position of the legend of the unit vector m ------------------------

\coordinate (rm) at (
{\s*\um*(\a*(1-\e^2)/(1+\e*cos(\fm)))*(cos(\O)*cos(\o+\fm)-cos(\i)*sin(\O)*sin(\o+\fm))},
{\s*\um*(\a*(1-\e^2)/(1+\e*cos(\fm)))*(sin(\O)*cos(\o+\fm)+cos(\i)*cos(\O)*sin(\o+\fm))},
{\s*\um*(\a*(1-\e^2)/(1+\e*cos(\fm)))*(sin(\i)*sin(\o+\fm))}
);

%---------------------------------- Position of the unit vector J ---------------------------------

\coordinate (J) at (
{\uJ*sin(\i)*sin(\O)},
{-\uJ*sin(\i)*cos(\O)},
{\uJ*cos(\i)}
);

%---------------------- Position of the legend of the unit vector J ------------------------

\coordinate (rJ) at (
{\s*\uJ*sin(\i)*sin(\O)},
{-\s*\uJ*sin(\i)*cos(\O)},
{\s*\uJ*cos(\i)}
);

%------------------Position of the rotation of the spin angular momentum -------------------------

\coordinate (sJ) at (
{(\uJ-0.1)*cos(\i)*cos(\O)},
{-(\uJ-0.1)*sin(\i)*cos(\O)},
{(\uJ-0.1)*cos(\i)}
);

%--------------------- Drawing the fundamental plane below and behind the orbital plane-----------------

\draw[white,opacity=0.5,fill=gray!50] plot[variable=\f,domain=\O:\O+180,samples=50]
({(\apl*(1-\epl^2)/(1+\epl*cos(\f)))*(cos(\Opl)*cos(\opl+\f)-cos(\ipl)*sin(\Opl)*sin(\opl+\f))},
{(\apl*(1-\epl^2)/(1+\epl*cos(\f)))*(sin(\Opl)*cos(\opl+\f)+cos(\ipl)*cos(\Opl)*sin(\opl+\f))},
{(\apl*(1-\epl^2)/(1+\epl*cos(\f)))*(sin(\ipl)*sin(\opl+\f))});

\draw[thick] plot[variable=\f,domain=\O:\O+98,samples=50]
({(\apl*(1-\epl^2)/(1+\epl*cos(\f)))*(cos(\Opl)*cos(\opl+\f)-cos(\ipl)*sin(\Opl)*sin(\opl+\f))},
{(\apl*(1-\epl^2)/(1+\epl*cos(\f)))*(sin(\Opl)*cos(\opl+\f)+cos(\ipl)*cos(\Opl)*sin(\opl+\f))},
{(\apl*(1-\epl^2)/(1+\epl*cos(\f)))*(sin(\ipl)*sin(\opl+\f))});

\draw[thick,dotted] plot[variable=\f,domain=\O+98:\O+180,samples=50]
({(\apl*(1-\epl^2)/(1+\epl*cos(\f)))*(cos(\Opl)*cos(\opl+\f)-cos(\ipl)*sin(\Opl)*sin(\opl+\f))},
{(\apl*(1-\epl^2)/(1+\epl*cos(\f)))*(sin(\Opl)*cos(\opl+\f)+cos(\ipl)*cos(\Opl)*sin(\opl+\f))},
{(\apl*(1-\epl^2)/(1+\epl*cos(\f)))*(sin(\ipl)*sin(\opl+\f))});

%----------------Drawing the part of the y axis below the orbital plane --------------

\draw[very thick,dotted] (O) -- (0,0.70,0);

%-------------------------Drawing the ellipse ----------------------------

\draw[white,opacity=0.75,fill=red!5] plot[variable=\x,domain=0:360,samples=50]
({(\a*(1-\e^2)/(1+\e*cos(\x)))*(cos(\O)*cos(\o+\x)-cos(\i)*sin(\O)*sin(\o+\x))},
{(\a*(1-\e^2)/(1+\e*cos(\x)))*(sin(\O)*cos(\o+\x)+cos(\i)*cos(\O)*sin(\o+\x))},
{(\a*(1-\e^2)/(1+\e*cos(\x)))*(sin(\i)*sin(\o+\x))});

\draw[very thick,red!60,dotted] plot[variable=\x,domain=\fdn:\fdn+43,samples=50]
({(\a*(1-\e^2)/(1+\e*cos(\x)))*(cos(\O)*cos(\o+\x)-cos(\i)*sin(\O)*sin(\o+\x))},
{(\a*(1-\e^2)/(1+\e*cos(\x)))*(sin(\O)*cos(\o+\x)+cos(\i)*cos(\O)*sin(\o+\x))},
{(\a*(1-\e^2)/(1+\e*cos(\x)))*(sin(\i)*sin(\o+\x))});

\draw[very thick,red!60,->] plot[variable=\x,domain=\fdn+43:\fan,samples=50]
({(\a*(1-\e^2)/(1+\e*cos(\x)))*(cos(\O)*cos(\o+\x)-cos(\i)*sin(\O)*sin(\o+\x))},
{(\a*(1-\e^2)/(1+\e*cos(\x)))*(sin(\O)*cos(\o+\x)+cos(\i)*cos(\O)*sin(\o+\x))},
{(\a*(1-\e^2)/(1+\e*cos(\x)))*(sin(\i)*sin(\o+\x))});

\draw[very thick,red!60] plot[variable=\x,domain=\fan:360,samples=50]
({(\a*(1-\e^2)/(1+\e*cos(\x)))*(cos(\O)*cos(\o+\x)-cos(\i)*sin(\O)*sin(\o+\x))},
{(\a*(1-\e^2)/(1+\e*cos(\x)))*(sin(\O)*cos(\o+\x)+cos(\i)*cos(\O)*sin(\o+\x))},
{(\a*(1-\e^2)/(1+\e*cos(\x)))*(sin(\i)*sin(\o+\x))});

\draw[very thick,red!60,->] plot[variable=\x,domain=0:\fdn,samples=50]
({(\a*(1-\e^2)/(1+\e*cos(\x)))*(cos(\O)*cos(\o+\x)-cos(\i)*sin(\O)*sin(\o+\x))},
{(\a*(1-\e^2)/(1+\e*cos(\x)))*(sin(\O)*cos(\o+\x)+cos(\i)*cos(\O)*sin(\o+\x))},
{(\a*(1-\e^2)/(1+\e*cos(\x)))*(sin(\i)*sin(\o+\x))});

%--------------------- Drawing the fundamental plane above the orbital plane-----------------

\draw[white,opacity=0.5,fill=gray!50] plot[variable=\f,domain=\O+180:\O+360,samples=50]
({(\apl*(1-\epl^2)/(1+\epl*cos(\f)))*(cos(\Opl)*cos(\opl+\f)-cos(\ipl)*sin(\Opl)*sin(\opl+\f))},
{(\apl*(1-\epl^2)/(1+\epl*cos(\f)))*(sin(\Opl)*cos(\opl+\f)+cos(\ipl)*cos(\Opl)*sin(\opl+\f))},
{(\apl*(1-\epl^2)/(1+\epl*cos(\f)))*(sin(\ipl)*sin(\opl+\f))});

\draw[thick] plot[variable=\f,domain=\O+180:\O+360,samples=50]
({(\apl*(1-\epl^2)/(1+\epl*cos(\f)))*(cos(\Opl)*cos(\opl+\f)-cos(\ipl)*sin(\Opl)*sin(\opl+\f))},
{(\apl*(1-\epl^2)/(1+\epl*cos(\f)))*(sin(\Opl)*cos(\opl+\f)+cos(\ipl)*cos(\Opl)*sin(\opl+\f))},
{(\apl*(1-\epl^2)/(1+\epl*cos(\f)))*(sin(\ipl)*sin(\opl+\f))})  node at (0.05,0.85,-0.27) {Fundamental plane};

%-----------------------Placing the x, y and z coordinate axes --------------------------

\draw[very thick,->] (O) -- (1.0,0,0)  node[anchor=north east] {$x$};

\draw[very thick,->] (0,0.70,0) -- (0,1.0,0)  node[anchor=north west] {$y$};

\draw[very thick,->] (O) -- (0,0,1.0)  node[anchor=south] {$z$};

%------------------------Placing the position vector in the orbital plane -----------------------------

\draw[very thick, bole,->] (O) -- (P) node at (rP) {$\boldsymbol{r}\ton{t}$};

%-------------------Placing the line of nodes in the fundamental plane ----------------------

\draw[very thick,cyan,->] (O) -- (AN) node at (rasc) {$\ascnode$}  node at ({\apl*cos(\O)},{\apl*sin(\O)},0.1) {$\boldsymbol{\hat{l}}$}  node at (-0.4,-0.55,-0.1) {Line of nodes} node at (rdesc) {$\descnode$};

\draw[very thick,cyan,dashed] (AN) -- ({\apl*cos(\O)},{\apl*sin(\O)},0);

\draw[very thick,dashed,cyan] (O) -- ({-\apl*cos(\O)},{-\apl*sin(\O)},0);

%---------------------- Placing the unit vector h orthogonal to the orbital plane --------------------

\draw[very thick,uclagold,->] (O) -- (h) node at (rh) {$\boldsymbol{\hat{h}}$};

%-------------------Placing the unit vector m in the orbital plane ---------------------------------

\draw[very thick,cyan,->] (O) -- (m) node at (rm) {$\boldsymbol{\hat{m}}$};

%-------------------------- Placing the longitude of the ascending node --------------------------------

\tdplotdrawarc[very thick,cyan,->]{(O)}{.33}{0}{\O}{anchor=north}{$\Omega$}

%------------------------------ Placing the inclination I------------------------------------------------------

\tdplotsetrotatedcoords{-90-\O}{-90}{0}
\tdplotdrawarc[very thick,uclagold,tdplot_rotated_coords,->]{(O)}{.75}{0}{\i}{anchor=south}{$I$}

%----------------------------Placing the argument of pericenter and the true anomaly --------------------------

\tdplotsetrotatedcoords{\O}{0}{\i}
\begin{scope}[tdplot_rotated_coords]
 \tdplotdrawarc[very thick,bole,->]{(O)}{.33}{0}{\o+\fo}{anchor=west}{$u(t)$}
\end{scope}

%------------------------ Placing the spin angular momentum J of the primary--------------

\draw[very thick,orange!95,->] (O) -- (0.45*0.353553,-0.45*0.353553,0.45*0.866025) node [above] {$\boldsymbol{\hat{J}}$};
\draw[very thick,orange!95,->] (O) -- (0.45*0.353553,-0.45*0.353553,0.45*0.866025) node at (0.37*0.353553,-0.37*0.353553,0.37*0.866025) {\AxisRotator[rotate = 300]};

%------------------------------Placing the RTN unit vectors ------------------------------------

%\draw [ao(english),very thick,->] (P) -- (R) node [right] {$\boldsymbol{\widehat{\mathcal{R}}}$};
%\draw [ao(english),very thick,->] (P) -- (T) node [above] {$\boldsymbol{\widehat{\mathcal{T}}}$};
%\draw [ao(english),very thick,->] (P) -- (H) node [above] {$\boldsymbol{\widehat{\mathcal{N}}}$};

%--------------------------Placing the test particle and the primary----------------------------------------------------

\shade[ball color=black] (P) circle (.1ex);

\shade[ball color=black] (O) circle (.22ex) node at (0,-0.1,-0.01) {$M$};

%--------------------------------------------------------------------------------------

\end{tikzpicture}
\caption[Prograde circular equatorial orbit]{Prograde circular equatorial orbit arbitrarily oriented in space with, say, $I = 30^\circ,\,\Omega = 45^\circ$. The orbital plane is aligned with the equator of the central body, and the test particle moves along the \textit{same} sense of rotation of the latter, so that $\boldsymbol{\hat{J}}\boldsymbol\cdot\boldsymbol{\hat{h}} = +1$.}\label{fig:pro}
\end{figure}
%%%%%%%%%%%%%%%%%%%%%%%%%%%%%%%%%%%%%%%%%%%%%%%%%%%%%%%%%%%%%%%%%%%%%%%%%%%%%%%%%%%%%%%%%%%%%%%%%%%%%%%%%%
%%%%%%%%%%%%%%%%%%%%%%%%%%%%%%%%%%%%%%%%%%%%%%%%%%%%%%%%%%%%%%%%%%%%%%%%%%%%%%%%%%%%%%%%%%%%%%%%%%%%%%%%%%

%%%%%%%%%%%%%%%%%%%%%%%%%%%%%%%%%%%%%%%%%%%%%%%%%%%%%%%%%%%%%%%%%%%%%%%%%%%%%%%%%%%%%%%%%%%%%%%%%%%%%%%%%%
%%%%%%%%%%%%%%%%%%%%%%%%%%%%%%%%%%%%%%%%%%%%%%%%%%%%%%%%%%%%%%%%%%%%%%%%%%%%%%%%%%%%%%%%%%%%%%%%%%%%%%%%%%
\tdplotsetmaincoords{70}{110}
\tdseteulerxyz
\begin{figure}
\centering
\begin{tikzpicture}[tdplot_main_coords,scale=7]
  \pgfmathsetmacro{\RA}{150} % RA of the spin axis
  \pgfmathsetmacro{\DEC}{75} % DEC of the spin axis
  \pgfmathsetmacro{\a}{1} % semimajor axis
  \pgfmathsetmacro{\e}{0.0} % eccentricity
  \pgfmathsetmacro{\i}{30} % inclination [deg]
  \pgfmathsetmacro{\O}{45} % right ascension of ascending node [deg]
  \pgfmathsetmacro{\o}{50} % argument of pericenter [deg]
  \pgfmathsetmacro{\fo}{60} % true anomaly [deg]
  \pgfmathsetmacro{\fpa}{0} % true anomaly at periapsis [deg]
  \pgfmathsetmacro{\faa}{180} % true anomaly at apoapsis [deg]
  \pgfmathsetmacro{\fan}{360-\o} % true anomaly at ascending node [deg]
  \pgfmathsetmacro{\fdn}{180-\o} % true anomaly at descending node [deg]
  \pgfmathsetmacro{\fm}{90-\o} % true anomaly at m [deg]
  \pgfmathsetmacro{\apl}{1} % semimajor axis of the fundamental plane
  \pgfmathsetmacro{\epl}{0} % eccentricity of the fundamental plane
  \pgfmathsetmacro{\ipl}{0} % inclination of the fundamental plane [deg]
  \pgfmathsetmacro{\Opl}{0} % right ascension of ascending node of the fundamental plane [deg]
  \pgfmathsetmacro{\opl}{0} % argument of pericenter of the fundamental plane [deg]
  \pgfmathsetmacro{\scl}{0.75} % scaling parameter for the dotted part of line of apsides below the fundamental plane
  \pgfmathsetmacro{\s}{1.1} % scaling parameter for legends
  \pgfmathsetmacro{\ofs}{0.03} % offset parameter for legends
  \pgfmathsetmacro{\u}{1} % offset parameter for the unit vectors
   \pgfmathsetmacro{\uJ}{0.35} % offset parameter for the unit vector J
   \pgfmathsetmacro{\usJ}{0.27} % offset parameter for the unit vector J
  \pgfmathsetmacro{\k}{0.3} % offset parameter for the RTN unit vectors
   \pgfmathsetmacro{\um}{0.8} % offset parameter for the m unit vector

%---------------------Position of the origin of the coordinates --------------------------------

\coordinate (O) at (0,0,0);

%-------------------Position of the test particle ------------------------------------

\coordinate (P) at (
{(\a*(1-\e^2)/(1+\e*cos(\fo)))*(cos(\O)*cos(\o+\fo)-cos(\i)*sin(\O)*sin(\o+\fo))},
{(\a*(1-\e^2)/(1+\e*cos(\fo)))*(sin(\O)*cos(\o+\fo)+cos(\i)*cos(\O)*sin(\o+\fo))},
{(\a*(1-\e^2)/(1+\e*cos(\fo)))*(sin(\i)*sin(\o+\fo))}
);

%------------------------- Position of the legend of the test particle ----------------------

\coordinate (rP) at (
{\s*(\a*(1-\e^2)/(1+\e*cos(\fo)))*(cos(\O)*cos(\o+\fo)-cos(\i)*sin(\O)*sin(\o+\fo))},
{\s*(\a*(1-\e^2)/(1+\e*cos(\fo)))*(sin(\O)*cos(\o+\fo)+cos(\i)*cos(\O)*sin(\o+\fo))-0.25},
{\s*(\a*(1-\e^2)/(1+\e*cos(\fo)))*(sin(\i)*sin(\o+\fo))-0.2}
);

%------------------------------ Position of the ascending node ----------------------------------------

\coordinate (AN) at (
{(\a*(1-\e^2)/(1+\e*cos(\fan)))*(cos(\O)*cos(\o+\fan)-cos(\i)*sin(\O)*sin(\o+\fan))},
{(\a*(1-\e^2)/(1+\e*cos(\fan)))*(sin(\O)*cos(\o+\fan)+cos(\i)*cos(\O)*sin(\o+\fan))},
{(\a*(1-\e^2)/(1+\e*cos(\fan)))*(sin(\i)*sin(\o+\fan))}
);

%-------------------------------- Position of the legend of the ascending node -----------------------

\coordinate (rAN) at (
{\s*(\a*(1-\e^2)/(1+\e*cos(\fan)))*(cos(\O)*cos(\o+\fan)-cos(\i)*sin(\O)*sin(\o+\fan))},
{\s*(\a*(1-\e^2)/(1+\e*cos(\fan)))*(sin(\O)*cos(\o+\fan)+cos(\i)*cos(\O)*sin(\o+\fan))},
{\s*(\a*(1-\e^2)/(1+\e*cos(\fan)))*(sin(\i)*sin(\o+\fan))}
);

%-------------------------------- Position of the end of the line of nodes in the fundamental plane -----------------------

\coordinate (lAN) at (
{\apl*(\a*(1-\e^2)/(1+\e*cos(\fan)))*(cos(\O)*cos(\o+\fan)-cos(\i)*sin(\O)*sin(\o+\fan))},
{\apl*(\a*(1-\e^2)/(1+\e*cos(\fan)))*(sin(\O)*cos(\o+\fan)+cos(\i)*cos(\O)*sin(\o+\fan))},
{\apl*(\a*(1-\e^2)/(1+\e*cos(\fan)))*(sin(\i)*sin(\o+\fan))}
);

%------------------------------Position of the legend of the unit vector l ---------------------------

\coordinate (rl) at (
{-\s*(\a*(1-\e^2)/(1+\e*cos(\fan)))*(cos(\O)*cos(\o+\fan)-cos(\i)*sin(\O)*sin(\o+\fan))},
{-\s*(\a*(1-\e^2)/(1+\e*cos(\fan)))*(sin(\O)*cos(\o+\fan)+cos(\i)*cos(\O)*sin(\o+\fan))},
{-\s*(\a*(1-\e^2)/(1+\e*cos(\fan)))*(sin(\i)*sin(\o+\fan))-\ofs}
);

%----------------------- Position of the symbol of the ascending node -------------------------

\coordinate (rasc) at (
{\s*(\a*(1-\e^2)/(1+\e*cos(\fan)))*(cos(\O)*cos(\o+\fan)-cos(\i)*sin(\O)*sin(\o+\fan))},
{\s*(\a*(1-\e^2)/(1+\e*cos(\fan)))*(sin(\O)*cos(\o+\fan)+cos(\i)*cos(\O)*sin(\o+\fan))},
{\s*(\a*(1-\e^2)/(1+\e*cos(\fan)))*(sin(\i)*sin(\o+\fan))-\ofs}
);

%------------------------- Position of the legend of the line of nodes --------------------------------

\coordinate (lon) at (
{(\a*(1-\e^2)/(1+\e*cos(\fan)))*(cos(\O)*cos(\o+\fan)-cos(\i)*sin(\O)*sin(\o+\fan))},
{1.5*\s*(\a*(1-\e^2)/(1+\e*cos(\fan)))*(sin(\O)*cos(\o+\fan)+cos(\i)*cos(\O)*sin(\o+\fan))},
{(\a*(1-\e^2)/(1+\e*cos(\fan)))*(sin(\i)*sin(\o+\fan))-\ofs}
);

%-------------------------- Position of the descending node ---------------------------------

\coordinate (DN) at (
{(\a*(1-\e^2)/(1+\e*cos(\fdn)))*(cos(\O)*cos(\o+\fdn)-cos(\i)*sin(\O)*sin(\o+\fdn))},
{(\a*(1-\e^2)/(1+\e*cos(\fdn)))*(sin(\O)*cos(\o+\fdn)+cos(\i)*cos(\O)*sin(\o+\fdn))},
{(\a*(1-\e^2)/(1+\e*cos(\fdn)))*(sin(\i)*sin(\o+\fdn))}
);

%------------------------ Position of the legend of the descending node--------------------------------

\coordinate (rDN) at (
{\s*(\a*(1-\e^2)/(1+\e*cos(\fdn)))*(cos(\O)*cos(\o+\fdn)-cos(\i)*sin(\O)*sin(\o+\fdn))},
{\s*(\a*(1-\e^2)/(1+\e*cos(\fdn)))*(sin(\O)*cos(\o+\fdn)+cos(\i)*cos(\O)*sin(\o+\fdn))},
{\s*(\a*(1-\e^2)/(1+\e*cos(\fdn)))*(sin(\i)*sin(\o+\fdn))}
);

%---------------------- Position of the symbol of the descending node------------------------

\coordinate (rdesc) at (
{\s*(\a*(1-\e^2)/(1+\e*cos(\fdn)))*(cos(\O)*cos(\o+\fdn)-cos(\i)*sin(\O)*sin(\o+\fdn))},
{\s*(\a*(1-\e^2)/(1+\e*cos(\fdn)))*(sin(\O)*cos(\o+\fdn)+cos(\i)*cos(\O)*sin(\o+\fdn))},
{\s*(\a*(1-\e^2)/(1+\e*cos(\fdn)))*(sin(\i)*sin(\o+\fdn))+\ofs}
);

%---------------------------------- Position of the unit vector h ---------------------------------

\coordinate (h) at (
{-\u*sin(\i)*sin(\O)},
{\u*sin(\i)*cos(\O)},
{-\u*cos(\i)}
);

%---------------------------------- Part of the orbital angular momentum below the orbital plane ---------------------------------

\coordinate (H) at (
{-0.58*sin(\i)*sin(\O)},
{0.58*sin(\i)*cos(\O)},
{-0.58*cos(\i)}
);

%---------------------- Position of the legend of the unit vector h ------------------------

\coordinate (rh) at (
{-\s*\u*sin(\i)*sin(\O)},
{\s*\u*sin(\i)*cos(\O)},
{-\s*\u*cos(\i)}
);

%-----------------------------------Position of the unit vector m -------------------------

\coordinate (m) at (
{\um*(\a*(1-\e^2)/(1+\e*cos(\fm)))*(cos(\O)*cos(\o+\fm)-cos(\i)*sin(\O)*sin(\o+\fm))},
{\um*(\a*(1-\e^2)/(1+\e*cos(\fm)))*(sin(\O)*cos(\o+\fm)+cos(\i)*cos(\O)*sin(\o+\fm))},
{\um*(\a*(1-\e^2)/(1+\e*cos(\fm)))*(sin(\i)*sin(\o+\fm))}
);

%------------------------------ Position of the legend of the unit vector m ------------------------

\coordinate (rm) at (
{\s*\um*(\a*(1-\e^2)/(1+\e*cos(\fm)))*(cos(\O)*cos(\o+\fm)-cos(\i)*sin(\O)*sin(\o+\fm))},
{\s*\um*(\a*(1-\e^2)/(1+\e*cos(\fm)))*(sin(\O)*cos(\o+\fm)+cos(\i)*cos(\O)*sin(\o+\fm))},
{\s*\um*(\a*(1-\e^2)/(1+\e*cos(\fm)))*(sin(\i)*sin(\o+\fm))}
);

%---------------------------------- Position of the unit vector J ---------------------------------

\coordinate (J) at (
{\uJ*sin(\i)*sin(\O)},
{-\uJ*sin(\i)*cos(\O)},
{\uJ*cos(\i)}
);

%---------------------- Position of the legend of the unit vector J ------------------------

\coordinate (rJ) at (
{\s*\uJ*sin(\i)*sin(\O)},
{-\s*\uJ*sin(\i)*cos(\O)},
{\s*\uJ*cos(\i)}
);

%------------------Position of the rotation of the spin angular momentum -------------------------

\coordinate (sJ) at (
{(\uJ-0.1)*cos(\i)*cos(\O)},
{-(\uJ-0.1)*sin(\i)*cos(\O)},
{(\uJ-0.1)*cos(\i)}
);

%--------------------- Drawing the fundamental plane below and behind the orbital plane-----------------

\draw[white,opacity=0.5,fill=gray!50] plot[variable=\f,domain=\O:\O+180,samples=50]
({(\apl*(1-\epl^2)/(1+\epl*cos(\f)))*(cos(\Opl)*cos(\opl+\f)-cos(\ipl)*sin(\Opl)*sin(\opl+\f))},
{(\apl*(1-\epl^2)/(1+\epl*cos(\f)))*(sin(\Opl)*cos(\opl+\f)+cos(\ipl)*cos(\Opl)*sin(\opl+\f))},
{(\apl*(1-\epl^2)/(1+\epl*cos(\f)))*(sin(\ipl)*sin(\opl+\f))});

\draw[thick] plot[variable=\f,domain=\O:\O+98,samples=50]
({(\apl*(1-\epl^2)/(1+\epl*cos(\f)))*(cos(\Opl)*cos(\opl+\f)-cos(\ipl)*sin(\Opl)*sin(\opl+\f))},
{(\apl*(1-\epl^2)/(1+\epl*cos(\f)))*(sin(\Opl)*cos(\opl+\f)+cos(\ipl)*cos(\Opl)*sin(\opl+\f))},
{(\apl*(1-\epl^2)/(1+\epl*cos(\f)))*(sin(\ipl)*sin(\opl+\f))});

\draw[thick,dotted] plot[variable=\f,domain=\O+98:\O+180,samples=50]
({(\apl*(1-\epl^2)/(1+\epl*cos(\f)))*(cos(\Opl)*cos(\opl+\f)-cos(\ipl)*sin(\Opl)*sin(\opl+\f))},
{(\apl*(1-\epl^2)/(1+\epl*cos(\f)))*(sin(\Opl)*cos(\opl+\f)+cos(\ipl)*cos(\Opl)*sin(\opl+\f))},
{(\apl*(1-\epl^2)/(1+\epl*cos(\f)))*(sin(\ipl)*sin(\opl+\f))});

%----------------Drawing the part of the y axis below the orbital plane --------------

\draw[very thick,dotted] (O) -- (0,0.70,0);

%-------------------------Drawing the ellipse ----------------------------

\draw[white,opacity=0.75,fill=red!5] plot[variable=\x,domain=0:360,samples=50]
({(\a*(1-\e^2)/(1+\e*cos(\x)))*(cos(\O)*cos(\o+\x)-cos(\i)*sin(\O)*sin(\o+\x))},
{(\a*(1-\e^2)/(1+\e*cos(\x)))*(sin(\O)*cos(\o+\x)+cos(\i)*cos(\O)*sin(\o+\x))},
{(\a*(1-\e^2)/(1+\e*cos(\x)))*(sin(\i)*sin(\o+\x))});

\draw[very thick,red!60,dotted,->] plot[variable=\x,domain=\fdn+43:\fdn,samples=50]
({(\a*(1-\e^2)/(1+\e*cos(\x)))*(cos(\O)*cos(\o+\x)-cos(\i)*sin(\O)*sin(\o+\x))},
{(\a*(1-\e^2)/(1+\e*cos(\x)))*(sin(\O)*cos(\o+\x)+cos(\i)*cos(\O)*sin(\o+\x))},
{(\a*(1-\e^2)/(1+\e*cos(\x)))*(sin(\i)*sin(\o+\x))});

\draw[very thick,red!60] plot[variable=\x,domain=\fdn+43:\fan,samples=50]
({(\a*(1-\e^2)/(1+\e*cos(\x)))*(cos(\O)*cos(\o+\x)-cos(\i)*sin(\O)*sin(\o+\x))},
{(\a*(1-\e^2)/(1+\e*cos(\x)))*(sin(\O)*cos(\o+\x)+cos(\i)*cos(\O)*sin(\o+\x))},
{(\a*(1-\e^2)/(1+\e*cos(\x)))*(sin(\i)*sin(\o+\x))});

\draw[very thick,red!60,->] plot[variable=\x,domain=360:\fan,samples=50]
({(\a*(1-\e^2)/(1+\e*cos(\x)))*(cos(\O)*cos(\o+\x)-cos(\i)*sin(\O)*sin(\o+\x))},
{(\a*(1-\e^2)/(1+\e*cos(\x)))*(sin(\O)*cos(\o+\x)+cos(\i)*cos(\O)*sin(\o+\x))},
{(\a*(1-\e^2)/(1+\e*cos(\x)))*(sin(\i)*sin(\o+\x))});

\draw[very thick,red!60] plot[variable=\x,domain=\fdn:0,samples=50]
({(\a*(1-\e^2)/(1+\e*cos(\x)))*(cos(\O)*cos(\o+\x)-cos(\i)*sin(\O)*sin(\o+\x))},
{(\a*(1-\e^2)/(1+\e*cos(\x)))*(sin(\O)*cos(\o+\x)+cos(\i)*cos(\O)*sin(\o+\x))},
{(\a*(1-\e^2)/(1+\e*cos(\x)))*(sin(\i)*sin(\o+\x))});

%--------------------- Drawing the fundamental plane above the orbital plane-----------------

\draw[white,opacity=0.5,fill=gray!50] plot[variable=\f,domain=\O+180:\O+360,samples=50]
({(\apl*(1-\epl^2)/(1+\epl*cos(\f)))*(cos(\Opl)*cos(\opl+\f)-cos(\ipl)*sin(\Opl)*sin(\opl+\f))},
{(\apl*(1-\epl^2)/(1+\epl*cos(\f)))*(sin(\Opl)*cos(\opl+\f)+cos(\ipl)*cos(\Opl)*sin(\opl+\f))},
{(\apl*(1-\epl^2)/(1+\epl*cos(\f)))*(sin(\ipl)*sin(\opl+\f))});

\draw[thick] plot[variable=\f,domain=\O+180:\O+360,samples=50]
({(\apl*(1-\epl^2)/(1+\epl*cos(\f)))*(cos(\Opl)*cos(\opl+\f)-cos(\ipl)*sin(\Opl)*sin(\opl+\f))},
{(\apl*(1-\epl^2)/(1+\epl*cos(\f)))*(sin(\Opl)*cos(\opl+\f)+cos(\ipl)*cos(\Opl)*sin(\opl+\f))},
{(\apl*(1-\epl^2)/(1+\epl*cos(\f)))*(sin(\ipl)*sin(\opl+\f))})  node at (0.05,0.85,-0.27) {Fundamental plane};

%-----------------------Placing the x, y and z coordinate axes --------------------------

\draw[very thick,->] (O) -- (1.0,0,0)  node[anchor=north east] {$x$};

\draw[very thick,->] (0,0.70,0) -- (0,1.0,0)  node[anchor=north west] {$y$};

\draw[very thick,->] (O) -- (0,0,1.0)  node[anchor=south] {$z$};

%------------------------Placing the position vector in the orbital plane -----------------------------

\draw[very thick, bole,->] (O) -- (P) node at (rP) {$\boldsymbol{r}\ton{t}$};

%-------------------Placing the line of nodes in the fundamental plane ----------------------

\draw[dashed,very thick,cyan] (O) -- (AN) node at (rasc) {$\descnode$} node at ({\apl*cos(\O+180)},{\apl*sin(\O+180)},-0.1) {$\boldsymbol{\hat{l}}$}  node at (0.5,0.6,0.1) {Line of nodes} node at (rdesc) {$\ascnode$};

\draw[very thick,cyan,->] (O) -- (DN);

%---------------------- Placing the unit vector h orthogonal to the orbital plane --------------------

\draw[dotted,very thick,uclagold] (O) -- (H);
\draw[very thick,uclagold,->] (H) -- (h) node at (rh) {$\boldsymbol{\hat{h}}$};

%-------------------Placing the unit vector m in the orbital plane ---------------------------------

\draw[very thick,cyan,->] (O) -- (m) node at (rm) {$\boldsymbol{\hat{m}}$};

%-------------------------- Placing the longitude of the ascending node --------------------------------

\tdplotdrawarc[very thick,cyan]{(O)}{.33}{0}{\O}{anchor=north}{$\Omega$}
\tdplotdrawarc[dotted,very thick,cyan,->]{(O)}{.33}{\O}{\O+180}{}{}

%------------------------------ Placing the inclination I------------------------------------------------------

\tdplotsetrotatedcoords{-90-\O}{-90}{0}
\tdplotdrawarc[very thick,uclagold,tdplot_rotated_coords]{(O)}{.75}{0}{-30}{}{}
\tdplotdrawarc[dotted,very thick,uclagold,tdplot_rotated_coords]{(O)}{.75}{-30}{-121}{}{}
\tdplotdrawarc[very thick,uclagold,tdplot_rotated_coords,->]{(O)}{.75}{-121}{\i-180}{anchor=north west}{$I$}

%----------------------------Placing the argument of pericenter and the true anomaly --------------------------

\tdplotsetrotatedcoords{\O+180}{0}{180-\i}
\tdplotdrawarc[very thick,bole,tdplot_rotated_coords,->]{(O)}{.33}{0}{180-\o-\fo}{anchor=south west}{$u(t)$}

%------------------------ Placing the spin angular momentum J of the primary--------------

\draw[very thick,orange!95,->] (O) -- (0.45*0.353553,-0.45*0.353553,0.45*0.866025) node [above] {$\boldsymbol{\hat{J}}$};
\draw[very thick,orange!95,->] (O) -- (0.45*0.353553,-0.45*0.353553,0.45*0.866025) node at (0.37*0.353553,-0.37*0.353553,0.37*0.866025) {\AxisRotator[rotate = 300]};

%------------------------------Placing the RTN unit vectors ------------------------------------

%\draw [ao(english),very thick,->] (P) -- (R) node [right] {$\boldsymbol{\widehat{\mathcal{R}}}$};
%\draw [ao(english),very thick,->] (P) -- (T) node [above] {$\boldsymbol{\widehat{\mathcal{T}}}$};
%\draw [ao(english),very thick,->] (P) -- (H) node [above] {$\boldsymbol{\widehat{\mathcal{N}}}$};

%--------------------------Placing the test particle and the primary----------------------------------------------------

\shade[ball color=black] (P) circle (.1ex);

\shade[ball color=black] (O) circle (.22ex) node at (0,-0.1,-0.01) {$M$};

%--------------------------------------------------------------------------------------

\end{tikzpicture}
\caption[Retrograde circular equatorial orbit]{Retrograde circular equatorial orbit arbitrarily oriented in space with, say, $I = 150^\circ,\,\Omega=225^\circ$. The orbital plane is aligned with the equator of the central body, and the test particle moves along the \textit{opposite} sense of rotation of the latter, so that $\boldsymbol{\hat{J}}\boldsymbol\cdot\boldsymbol{\hat{h}} = -1$.}\label{fig:retro}
\end{figure}
%%%%%%%%%%%%%%%%%%%%%%%%%%%%%%%%%%%%%%%%%%%%%%%%%%%%%%%%%%%%%%%%%%%%%%%%%%%%%%%%%%%%%%%%%%%%%%%%%%%%%%%%%%
%%%%%%%%%%%%%%%%%%%%%%%%%%%%%%%%%%%%%%%%%%%%%%%%%%%%%%%%%%%%%%%%%%%%%%%%%%%%%%%%%%%%%%%%%%%%%%%%%%%%%%%%%%
%
In this case, according to \rfrs{Irate}{Orate}, the satellite's orbital plane stays unchanged in space, being aligned with  the equator of the primary. The same occurs also if the latter is oblate since the resulting Newtonian precessions of the node and the inclination  are proportional  to $\boldsymbol{\hat{J}}\boldsymbol{\cdot}\boldsymbol{\hat{m}}$ and $\boldsymbol{\hat{J}}\boldsymbol{\cdot}\boldsymbol{\hat{l}}$, respectively \citep{2017EPJC...77..439I}. In particular, the line of nodes remains fixed, and $\boldsymbol{\hat{l}}$ is constant; thus, it can be naturally assumed as a reference direction in the orbital plane.
Instead, in general, the line of the apsides changes because of \rfr{orate}.
The argument of latitude $u$, counted just from $\boldsymbol{\hat{l}}$ taken as reference polar axis, can be used as polar angle reckoning the instantaneous position of the test particle along its orbit. According to \rfrs{ARLT}{ANLT}, if \rfrs{kh}{klm} hold,  the normal component of the LT acceleration vanishes, and the gravitomagnetically perturbed motion is entirely in-plane.

If the orbit is also circular, then \rfr{ALT} is only radial, reducing to
\eqi
\mathrm{A}_\mathrm{LT} = \pm \rp{2\,\nk\,G\,J}{c^2\,r_0^2}.\lb{Arlt}
\eqf
The \virg{+} and \virg{-} signs in \rfr{Arlt} correspond to the prograde $\ton{\boldsymbol{\hat{J}}\boldsymbol{\cdot}\boldsymbol{\hat{h}} = +1}$ and retrograde $\ton{\boldsymbol{\hat{J}}\boldsymbol{\cdot}\boldsymbol{\hat{h}} = -1}$  motion with respect to the sense of rotation of the primary, respectively; Figure\,\ref{fig:pro} shows the first case, while the second one is depicted in Figure\,\ref{fig:retro}.

It can be noted that when the test particle orbits the central body along the \textit{same} sense of the rotation of the latter, i.e. for the \virg{+} sign,  \rfr{Arlt} is radially directly \textit{outward}, \textit{weakening} the overall gravitational pull to which the test particle is subjected;  thus, an \textit{increase} in the time required to complete a full orbital revolution, which, in this case, can be naturally assumed as the time interval between two consecutive crossings of the fixed line of the nodes, is expected since the motion is \textit{slower}. On the contrary, when the satellite moves in the \textit{opposite} direction with respect to the sense of rotation of its primary, the overall gravitational tug experienced by the former is \textit{enhanced} since \rfr{Arlt} is radially directed \textit{inward} ($\virg{-}$ sign), the motion is \textit{faster}, and the completion of an orbital revolution is expected to take \textit{less} time. \citet{MitPG} obtained, incorrectly, the opposite result, while \citet{Vladi87} agreed with the picture just outlined.

In the next Section, we will quantitatively assess such distinctive features of motion, whose non-Machian nature was discussed by \citet{Mashetal01}, by numerically integrating the equations of motion of a fictitious test particle orbiting a hypothetical massive, spinning body.
\section{Numerical integrations of the equations of motion}
\begin{figure}[ht!]
\centering
\centerline{
\vbox{
\begin{tabular}{c}
\epsfxsize= 12.5 cm\epsfbox{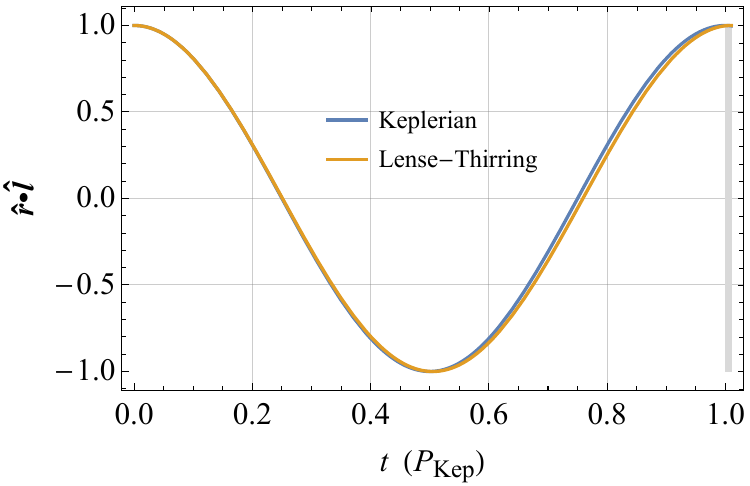} \\ \epsfxsize= 12.5 cm\epsfbox{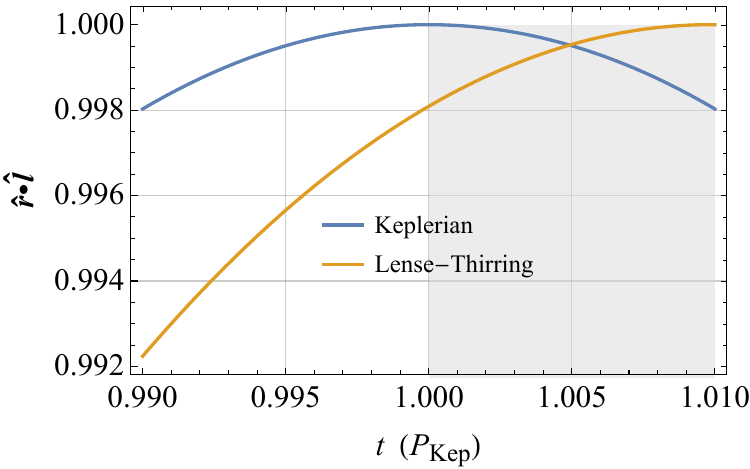}\\
\end{tabular}
}
}
\caption{
Plot of the numerically produced time series of $\boldsymbol{\hat{r}}\boldsymbol{\cdot}\boldsymbol{\hat{l}}$ versus time $t$ with and without the LT acceleration of \rfr{ALT} for a circular and equatorial orbit arbitrarily oriented in space starting from the ascending node with $\boldsymbol{\hat{J}}\boldsymbol{\cdot}\boldsymbol{\hat{h}} = +1$ (prograde motion). The relevant physical parameters of the putative primary are in Table\ref{tab1}. For the sake of clarity, a value of $J$ such that $\mathrm{A}_\mathrm{LT}/\mathrm{A}_\mathrm{N}=0.005$ was adopted. It turns out that the time intervals between two consecutive crossings of the line of the nodes, which is fixed both in the Keplerian and LT cases, differ by $ + 8\pi\,J/\ton{M\,c^2}$ which, in units of $P_\mathrm{Kep}$, amounts to $0.01$. The shaded area represents such a retardation with respect to the Keplerian period.
}\label{figure:1}
\end{figure}

\lb{sec:4}
Let us consider a fictitious test particle orbiting a massive, spinning body with the same mass and spin axis orientation of, say, Jupiter, whose relevant physical parameters are listed in Table\,\ref{tab1}.
\begin{table}[ht!]
\caption{Relevant physical parameters of Jupiter \citep{iers10,2018Natur.555..220I}.}\lb{tab1}
\begin{center}
\begin{tabular}{|l|l|l|}
  \hline
Parameter  & Units & Numerical value \\
\hline
$\mu$ & $\textrm{m}^3~\textrm{s}^{-2}$ & $1.26713\times 10^{17}$ \citep{iers10}\\
%
%
%$J_2$ & $\times 10^{-6}$ & $14696.572$ \citep{2018Natur.555..220I}\\
%
%
%$J$ & $\textrm{kg~m}^2$~$\textrm{s}^{-1}$ & $6.9\times 10^{38}$ \citep{2003AJ....126.2687S}\\
%
$\alpha_J$ & $\textrm{deg}$ & $268.057132$ \citep{2018Natur.555..220I}\\
$\delta_J$ & $\textrm{deg}$ & $64.497159$ \citep{2018Natur.555..220I}\\
%
%$R_\mathrm{e}$ & $\textrm{km}$ & $71492$ \citep{2007CeMDA..98..155S}\\
%
%
%
%
\hline
\end{tabular}
\end{center}
\end{table}
It is assumed that the satellite moves along a circular and equatorial prograde orbit starting from the ascending node along the line of the nodes, i.e.,  $\boldsymbol{\hat{J}}\boldsymbol{\cdot}\boldsymbol{\hat{h}} = +1$  and ${\boldsymbol{\hat{r}}}_0\boldsymbol{\cdot}\boldsymbol{\hat{l}} = +1$. Its equations of motion, in Cartesian  coordinates, are numerically integrated with and without the LT acceleration of \rfr{ALT} starting from the same initial conditions. The angular momentum $J$ of the fictitious primary is assumed large enough to easily visualize its gravitomagnetic effects remaining, at the same time, to a level compatible with the 1pN approximation. In particular, a value for $J$ is adopted such that the ratio of \rfr{Arlt} to the Newtonian monopole term is as little as $0.005$. Then, a time series for the cosine of the angle between the line of the nodes and the test particle's position vector, i.e. $\boldsymbol{\hat{r}}\boldsymbol{\cdot}\boldsymbol{\hat{l}}$, is numerically produced in both the runs and plotted versus time in order to reckon the instant of time when $\boldsymbol{\hat{l}}$ is crossed again after the initial instant, i.e., when $\boldsymbol{\hat{r}}\boldsymbol{\cdot}\boldsymbol{\hat{l}} = +1$ again. Figure\,\ref{figure:1} displays both the Keplerian and the gravitomagnetic time series over one orbital revolution.
It turns out that, when \rfr{ALT} is included in the equations of motion, the passage of the test particle at the line of the nodes occurs after an amount of time equal to the Keplerian orbital period \textit{increased} by the gravitomagnetic correction
\eqi
\delta T_\mathrm{gvm}^{+} := + 8\pi\,\rp{J}{M\,c^2},\lb{DPLT}
\eqf
which, in the fictional example considered, amounts to the $1\%$ of $P_\mathrm{Kep}$; such a \textit{retardation} is represented by the shaded area in Figure\,\ref{figure:1}.

\begin{figure}[ht!]
\centering
\centerline{
\vbox{
\begin{tabular}{c}
\epsfxsize= 12.5 cm\epsfbox{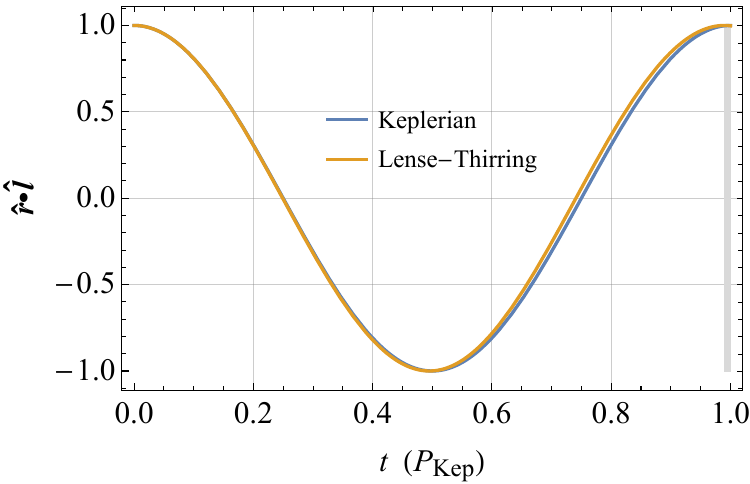} \\ \epsfxsize= 12.5 cm\epsfbox{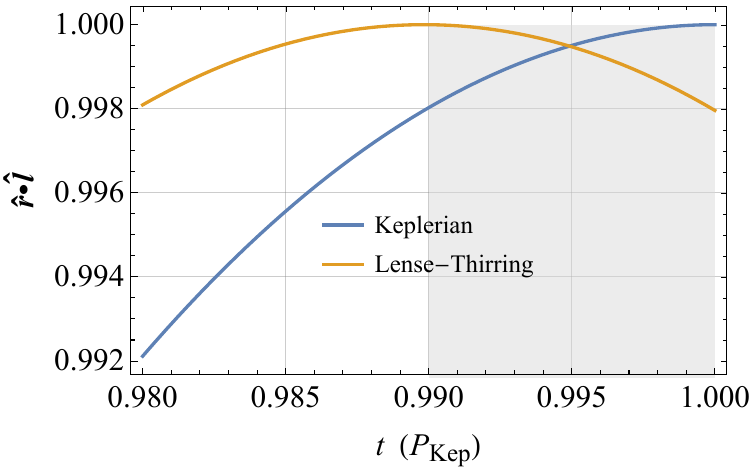}\\
\end{tabular}
}
}
\caption{
Plot of the numerically produced time series of $\boldsymbol{\hat{r}}\boldsymbol{\cdot}\boldsymbol{\hat{l}}$ versus time $t$ with and without the LT acceleration of \rfr{ALT} for a circular and equatorial orbit arbitrarily oriented in space starting from the ascending node with $\boldsymbol{\hat{J}}\boldsymbol{\cdot}\boldsymbol{\hat{h}} =  -1$ (retrograde motion). The relevant physical parameters of the putative primary are in Table\ref{tab1}. For the sake of clarity, a value of $J$ such that $\mathrm{A}_\mathrm{LT}/\mathrm{A}_\mathrm{N}=0.005$ was adopted. It turns out that the time intervals between two consecutive crossings of the line of the nodes, which is fixed both in the Keplerian and LT cases, differ by $-8\pi\,J/\ton{M\,c^2}$ which, in units of $P_\mathrm{Kep}$, amounts to $-0.01$. The shaded area represents such an advance with respect to the Keplerian period.
}\label{figure:2}
\end{figure}

Figure\,\ref{figure:2} depicts the same scenario, but for a retrograde orbit, i.e., with $\boldsymbol{\hat{J}}\boldsymbol{\cdot}\boldsymbol{h} = -1$. In this case, the crossing of the line of the nodes occurs \textit{in advance} with respect to the purely Keplerian case by the same quantity of \rfr{DPLT}; such an \textit{advance} is represented by the shaded area in Figure\,\ref{figure:2}.
Further numerical integrations showed that by varying the orbital radius does not affect the magnitude of the gravitomagnetic retardation/advance.

\begin{figure}[ht!]
\centering
\centerline{
\vbox{
\begin{tabular}{c}
\epsfxsize= 12.5 cm\epsfbox{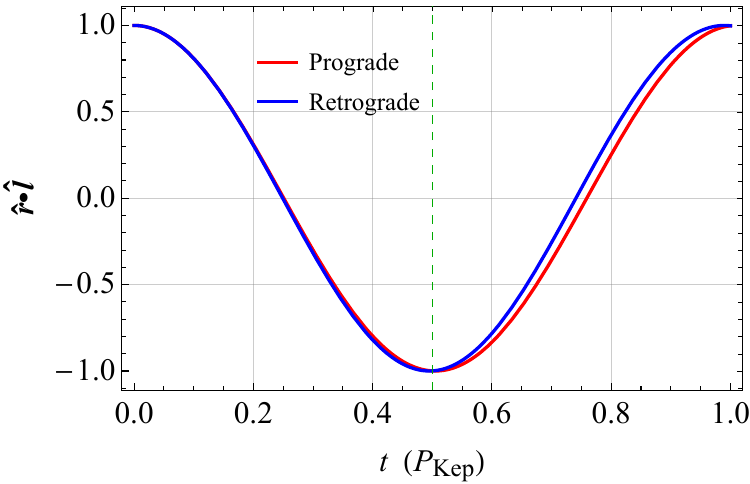} \\
\epsfxsize= 12.5 cm\epsfbox{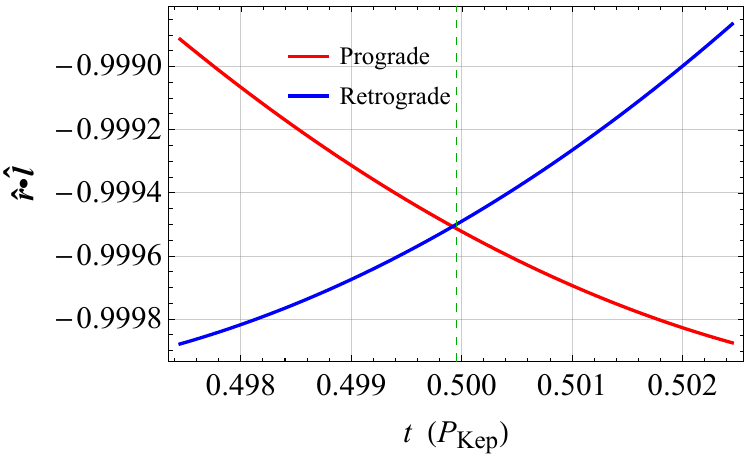}
\end{tabular}
}
}
\caption{
Plots of the numerically produced time series of  ${\boldsymbol{\hat{r}}}^\mathrm{+}\boldsymbol{\cdot}\boldsymbol{\hat{l}}$ and ${\boldsymbol{\hat{r}}}^\mathrm{-}\boldsymbol{\cdot}\boldsymbol{\hat{l}}$ versus time $t$ with the LT acceleration of \rfr{ALT} for a circular and equatorial orbit arbitrarily oriented in space starting from the ascending node with $\boldsymbol{\hat{J}}\boldsymbol{\cdot}\boldsymbol{\hat{h}} =  \pm 1$. The relevant physical parameters of the putative primary are in Table\ref{tab1}. For the sake of clarity, a value of $J$ such that $\mathrm{A}_\mathrm{LT}/\mathrm{A}_\mathrm{N}=0.005$ was adopted. It turns out that the two counter-orbiting test particles meet at $\overline{t} = P_\mathrm{Kep}/2\,\ton{1 - \epsilon_\mathrm{gvm}^2}$, with $\epsilon_\mathrm{gvm}$ given by \rfr{epsi}, represented by the vertical dashed green line.
}\label{figure:3}
\end{figure}

Figure\,\ref{figure:3} deals with the instant of time $\overline{t}$ when the two counter-orbiting test particles meet each other for the first time after their common start. In the Keplerian case, such an event would occur just after half a revolution at $\overline{\upvarphi}_\mathrm{Kep} = \pi$. Instead, when the LT acceleration of \rfr{ALT} is present, it is expected that the retrograde particle, which moves faster, reaches the prograde one, which is slower, \textit{before} than in the static case.
In Figure\,\ref{figure:3}, the numerically produced time series of ${\boldsymbol{\hat{r}}}^{\pm}\boldsymbol{\cdot}\boldsymbol{\hat{l}}$, obtained by including \rfr{ALT} in the equations of motion of both particles, are plotted together, and the time when they cross is marked with a vertical dashed green line. It turns out that
\eqi
\overline{t} = \rp{P_\mathrm{Kep}}{2}\,\ton{1 - \epsilon_\mathrm{gvm}^2},\lb{tast}
\eqf
with
\eqi
\epsilon_\mathrm{gvm} := \rp{\left|\delta T_\mathrm{gvm}\right|}{P_\mathrm{Kep}}= 8\pi\,\rp{J}{M\,c^2\,P_\mathrm{Kep}}.\lb{epsi}
\eqf
\section{Investigating the discrepancy with the standard scenario}\lb{sec:5}
Theory predicts just what we came up with in Section\,\ref{sec:4}, as shown by $\mathrm{Eq.}\,\ton{74}$ and $\mathrm{Eq.}\,\ton{77}$ of \citet{2016MNRAS.460.2445I}. Such an outcome has an impact also on the gravitomagnetic clock effect which, now, is
\eqi
\Delta T_\mathrm{gvm}:= T^{+}_\mathrm{gvm} - T_\mathrm{gvm}^{-} = 16\pi\,\rp{J}{M\,c^2}.\lb{mia}
\eqf
For the Earth, since its angular momentum per unit mass amounts to \citep{iers10} $J_\oplus/M_\oplus\simeq 9\times 10^8\,\mathrm{m}^2\,\mathrm{s}^{-1}$,
\rfr{mia} returns $\Delta T^\oplus_\mathrm{gvm} \simeq 5\times 10^{-7}\,\mathrm{s}$. The angular momentum of Jupiter is \citep{2003AJ....126.2687S} $J_\mathrm{Jup}\simeq 6.9\times 10^{38}\,\mathrm{kg\,m}^2\,\mathrm{s}^{-1}$; then, one has $\Delta T^\mathrm{Jup}_\mathrm{gvm} \simeq 2\times 10^{-4}\,\mathrm{s}$. Since the Sun's angular momentum is \citep{1998MNRAS.297L..76P} $J_\odot\simeq 1.90\times 10^{41}\,\mathrm{kg\,m}^2\,\mathrm{s}^{-1}$, the solar gravitomagnetic clock effect is $\Delta T^\odot_\mathrm{gvm} \simeq 5\times 10^{-5}\,\mathrm{s}$.
Interestingly, for a hypothetical pair of counter-revolving S-stars in the equatorial plane of the supermassive black hole in Sgr A$^\ast$ at the Galactic Center, \rfr{mia}, calculated for $J_\bullet =\chi_\bullet\,\ton{M_\bullet^2\,G}/c$ with \citep{2023MNRAS.tmp.3096D} $\chi_\bullet\simeq 0.90$ and \citep{2009ApJ...692.1075G} $M_\bullet\simeq 4.1\times 10^6\,M_\odot$, yields $\Delta T^\bullet_\mathrm{gvm} \simeq 9.13\times 10^{2}\,\mathrm{s}$.

Also the analytical calculation of $\overline{t}$, obtained by following the reasoning by \citet{Mashetal01} extended to the approach by  \citet{2016MNRAS.460.2445I}, agrees with \rfr{tast}. One has, first, to compute with the theoretical model used the instant of times $\overline{t}^{\,\,\pm}$ corresponding to the fixed angles $\overline{\upvarphi}$ and $2\pi-\overline{\upvarphi}$ for the prograde and retrograde directions, respectively. Then, by imposing the condition $\overline{t}^{\,\,+} = \overline{t}^{\,\,-} = \overline{t}$, one gets $\overline{t}$ and $\overline{\upvarphi}$.
By using the calculational scheme in \citet{2016MNRAS.460.2445I}, it can be obtained
\begin{align}
\overline{t}^{\,\,+} \lb{up}& =  \ton{\rp{1}{n_\mathrm{Kep}}  + 4\,\rp{J}{M\,c^2}}\,\overline{u}, \\ \nonumber \\
\overline{t}^{\,\,-} \lb{um}& = \ton{\rp{1}{n_\mathrm{Kep}}  - 4\,\rp{J}{M\,c^2}}\,\ton{2\pi - \overline{u}};
\end{align}
see below for further details.
By requiring that the right-hand sides of \rfrs{up}{um} are equal and solving for $\overline{u}$, one gets
\eqi
\overline{u} = \pi\,\ton{1 - \epsilon_\mathrm{gvm}}.\lb{uol}
\eqf
Inserting \rfr{uol} in any of \rfrs{up}{um} yields
\eqi
\overline{t} = \rp{P_\mathrm{Kep}}{2}\,\ton{1 - \epsilon_{gvm}^2},
\eqf
in agreement with \rfr{tast}.

On the other hand, as far as $\Delta T_\mathrm{gvm}$ is concerned, a discrepancy by a factor of $4$ occurs with respect to the so-far accepted  result\footnote{\citet{MitPG}, incorrectly, obtained the opposite signs.} \citep{Vladi87,1993PhLA..181..353C,Mash97,You98,Tart00,Mashetal99,MashSan00,Mashetal01}
\eqi
\delta T_\mathrm{gvm}^{\pm} := \pm 2\pi\,\rp{J}{M\,c^2}\lb{MitPG}
\eqf
which can be obtained by equating the centripetal acceleration ${\upomega}^2\,r_0$ to the sum of the Newtonian monopole plus \rfr{Arlt}. Indeed, from
\eqi
\upomega^2\,r_0 = \rp{\mu}{r_0^2} \mp\rp{2\,\nk\,G\,J}{c^2\,r_0^2}\lb{simple}
\eqf
one gets
\eqi
\dert\upvarphi t = \pm\nk\,\sqrt{1 \mp \rp{2\,\nk\,J}{M\,c^2}}
\eqf
which yields
\eqi
\dert t \upvarphi = \pm\rp{1}{\nk\,\sqrt{1\mp \rp{2\,\nk\,J}{M\,c^2}}}\simeq \pm\rp{1}{\nk}\,\ton{1\pm \rp{\nk\,J}{M\,c^2}}.\lb{dtdphi}
\eqf
By integrating \rfr{dtdphi} with respect to $\upvarphi$ from 0 to $+2\pi$ for the prograde motion and  from 0 to $-2\pi$ for the retrograde one,  the LT orbital period
\eqi
T^{\pm}_\mathrm{gvm} =  P_\mathrm{Kep} + \delta T_\mathrm{gvm}^{\pm}:= \rp{2\pi}{\nk}\pm 2\pi\,\rp{J}{M\,c^2}
\eqf
is obtained.
Note that $\upvarphi$ is a polar angle counted from some fixed reference polar axis in the orbital plane aimed to instantaneously locate the test particle along its circular orbit; thus, for an equatorial orbit, it is straightforward to identify the fixed line of the nodes with the reference direction and $\upvarphi$ with the argument of latitude $u$.

The explanation in the aforementioned discrepancy likely resides in the fact that the more general calculation in \citet{2016MNRAS.460.2445I}, made by using the non-singular elements $q$ and $k$, accounts for the fact that, during two consecutive crossings of the line of the nodes, the orbital elements in terms of which $\mathrm{d}t/\mathrm{d}u$ is parameterized, i.e. $p,\,q$ and $k$, do actually change instantaneously.
In the general case, also the line of the nodes does not stay fixed; such a feature is captured by the calculation in \citet{2016MNRAS.460.2445I} as well.
It turns out that such an effect does not vanish even in the limit $q,\,k\rightarrow 0$ corresponding to a circular orbit.
Indeed, a step-by-step analysis of the calculation in \citet{2016MNRAS.460.2445I} made with the LT acceleration of \rfr{ALT} shows that \rfr{DPLT} comes from the sum of
\begin{align}
\int_0^{2\pi}\derp{\ton{\mathrm{d}t/\mathrm{d}u}} q\,\Delta q\ton{u}\,\mathrm{d}u ,\\ \nonumber\\
\int_0^{2\pi}\derp{\ton{\mathrm{d}t/\mathrm{d}u}} k\,\Delta k\ton{u}\,\mathrm{d}u
\end{align}
which, for $q,\,k\rightarrow 0$, do not vanish yielding
\begin{align}
4\,\rp{J}{M\,c^2}\,\int_0^{2\pi}\cos u\,\ton{\cos u - \cos u_0}\,\mathrm{d}u \lb{kla}&= + 4\pi\,\rp{J}{M\,c^2},\\ \nonumber\\
4\,\rp{J}{M\,c^2}\,\int_0^{2\pi}\sin u\,\ton{\sin u - \sin u_0}\,\mathrm{d}u \lb{klu}&= + 4\pi\,\rp{J}{M\,c^2}.
\end{align}
Instead, it turns out that $\partial\ton{\mathrm{d}t/\mathrm{d}u}/\partial p\,\Delta p\ton{u} = 0$ since, in the limit $q,\,k\rightarrow 0$, the instantaneous variation $\Delta p\ton{u}$ of the semilatus rectum $p$ vanishes. Also the term due to the change of the line of the nodes containing $\mathrm{d}\Omega/\mathrm{d}t$ \citep{2016MNRAS.460.2445I} is zero for an equatorial orbit since, according to \rfr{ANLT} and \rfrs{kh}{klm},  $\mathrm{A}_N^\mathrm{LT} = 0$.
The opposite sign in \rfrs{kla}{klu} is obtained for the retrograde motion.
It can, now, be noted that \rfr{up} comes from the sum of
\begin{align}
\int_0^{\overline{u}}\derp{\ton{\mathrm{d}t/\mathrm{d}u}} q\,\Delta q\ton{u}\,\mathrm{d}u & = +2\,\rp{J}{M\,c^2}\,\overline{u}, \\ \nonumber \\
\int_0^{\overline{u}}\derp{\ton{\mathrm{d}t/\mathrm{d}u}} k\,\Delta k\ton{u}\,\mathrm{d}u & = +2\,\rp{J}{M\,c^2}\,\overline{u},
\end{align}
calculated with $\boldsymbol{\hat{J}}\boldsymbol{\cdot}\boldsymbol{\hat{h}} = +1$.
On the other hand, \rfr{um} is the sum of
\begin{align}
\int_0^{2\pi - \overline{u}}\derp{\ton{\mathrm{d}t/\mathrm{d}u}} q\,\Delta q\ton{u}\,\mathrm{d}u & = -2\,\rp{J}{M\,c^2}\,\ton{2\pi - \overline{u}}, \\ \nonumber \\
\int_0^{2\pi - \overline{u}}\derp{\ton{\mathrm{d}t/\mathrm{d}u}} k\,\Delta k\ton{u}\,\mathrm{d}u & = -2\,\rp{J}{M\,c^2}\,\ton{2\pi - \overline{u}},
\end{align}
calculated with $\boldsymbol{\hat{J}}\boldsymbol{\cdot}\boldsymbol{\hat{h}} = -1$.

Instead,
the integration based on \rfr{simple} is performed by considering only $\upvarphi$ as variable during an orbital revolution, all the rest being kept fixed.
\section{The apsidal period}
\begin{figure}[ht!]
\centering
\centerline{
\vbox{
\begin{tabular}{c}
\epsfxsize= 12.5 cm\epsfbox{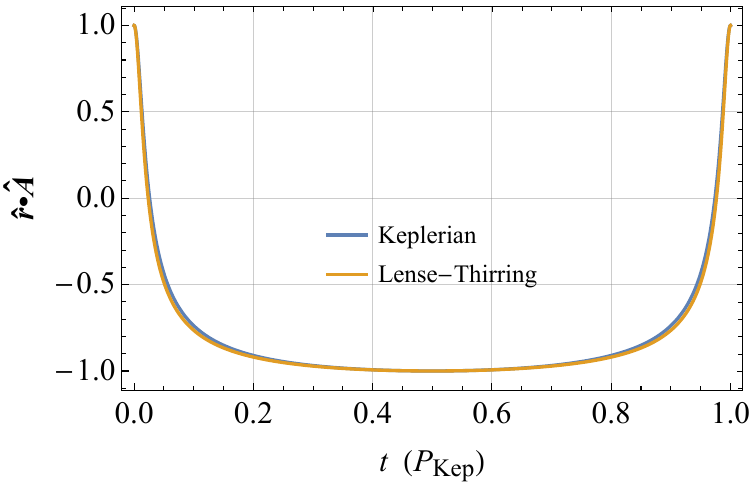} \\ \epsfxsize= 12.5 cm\epsfbox{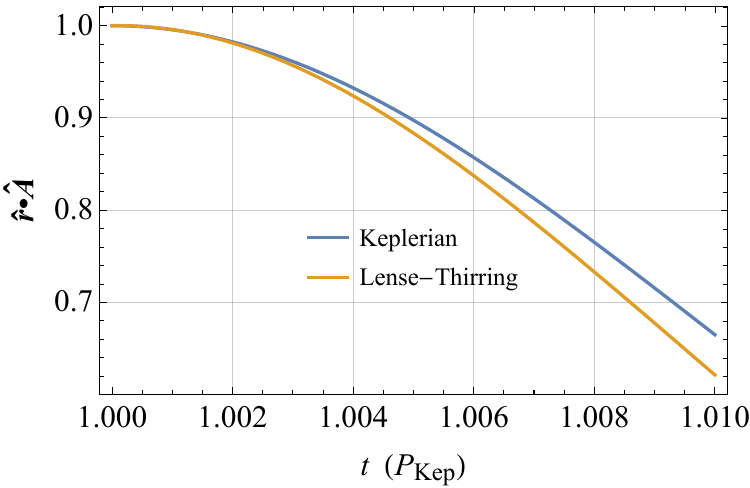}\\
\end{tabular}
}
}
\caption{
Plot of the numerically produced time series of $\boldsymbol{\hat{r}}\boldsymbol{\cdot}\boldsymbol{\hat{A}}$ versus time $t$ with and without the LT acceleration of \rfr{ALT} for an elliptical ($e=0.8$) and equatorial orbit arbitrarily oriented in space starting from the periapsis with $\boldsymbol{\hat{J}}\boldsymbol{\cdot}\boldsymbol{\hat{h}} = +1$ (prograde motion). The relevant physical parameters of the putative primary are in Table\ref{tab1}. For the sake of clarity, a value of $J$ such that $\mathrm{A}_\mathrm{LT}/\mathrm{A}_\mathrm{N}=0.005$ was adopted. It turns out that the time intervals between two consecutive crossings of the apsidal line, which varies when \rfr{ALT} is present, are identical in both the Keplerian and LT cases.
}\label{figure:4}
\end{figure}

\begin{figure}[ht!]
\centering
\centerline{
\vbox{
\begin{tabular}{c}
\epsfxsize= 12.5 cm\epsfbox{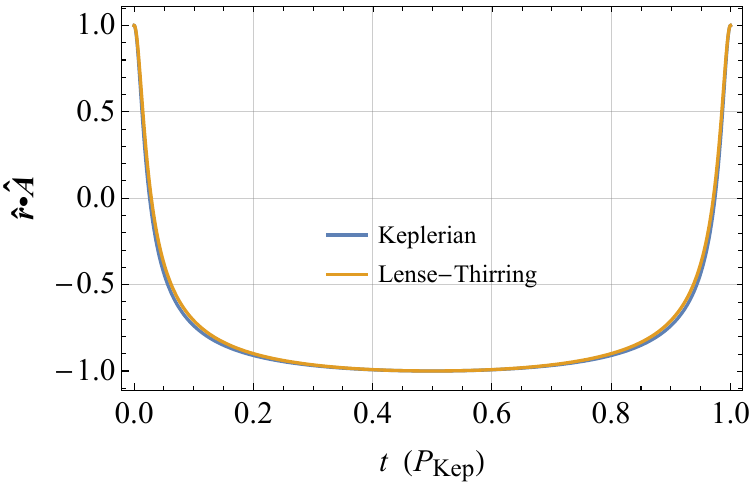} \\ \epsfxsize= 12.5 cm\epsfbox{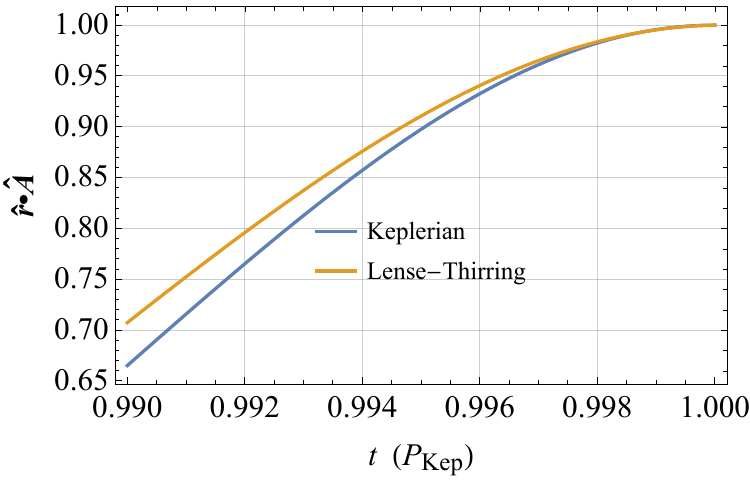}\\
\end{tabular}
}
}
\caption{
Plot of the numerically produced time series of $\boldsymbol{\hat{r}}\boldsymbol{\cdot}\boldsymbol{\hat{A}}$ versus time $t$ with and without the LT acceleration of \rfr{ALT} for an elliptical ($e=0.8$) and equatorial orbit arbitrarily oriented in space starting from the periapsis with $\boldsymbol{\hat{J}}\boldsymbol{\cdot}\boldsymbol{\hat{h}} = -1$ (retrograde motion). The relevant physical parameters of the putative primary are in Table\ref{tab1}. For the sake of clarity, a value of $J$ such that $\mathrm{A}_\mathrm{LT}/\mathrm{A}_\mathrm{N}=0.005$ was adopted. It turns out that the time intervals between two consecutive crossings of the apsidal line, which varies when \rfr{ALT} is present, are identical in both the Keplerian and LT cases.
}\label{figure:5}
\end{figure}

\lb{sec:6}
Our numerical tests confirm also another result of \citet{2016MNRAS.460.2445I}; in the case of an elliptical orbit, there is no LT correction to the apsidal period, i.e. the time interval between two consecutive crossings at the (moving) periapsis, for both the senses of motion. Figure\,\ref{figure:4} plots the numerically produced time series of the cosine of the angle between the position vector and the Laplace-Runge-Lenz vector \textbf{\textit{A}} over an orbital revolution with and without the LT acceleration of \rfr{ALT} for a highly elliptical and equatorial orbit arbitrarily oriented in space: both the runs share the same initial conditions corresponding to ${\boldsymbol{\hat{r}}}_0\boldsymbol{\cdot}{\boldsymbol{A}}_0 = +1$ and $\boldsymbol{\hat{J}}\boldsymbol{\cdot}\boldsymbol{\hat{h}} = +1$ (prograde motion along an equatorial orbit). It turns out that the apsidal periods are the same in both the Keplerian and LT cases.
The same occurs also for $\boldsymbol{\hat{J}}\boldsymbol{\cdot}\boldsymbol{\hat{h}} = -1$ (retrograde motion along an equatorial orbit), as shown by Figure\,\ref{figure:5}.
Further numerical integrations show that the anomalistic period is not impacted by the gravitomagnetic field of the primary also for non-equatorial orbits, in agreement with \citet{2016MNRAS.460.2445I}.

\textcolor{black}{The fact that the gravitomagnetic apsidal period is identical to the Keplerian one can be intuitively justified since, as per \rfrs{Irate}{Orate}, there is not net shift per orbit of the mean anomaly at epoch $\eta$. Indeed, from the definition of the mean anomaly
\eqi
\mathcal{M}\ton{t}:= \eta + n_\mathrm{Kep}\,\ton{t - t_0} = n_\mathrm{Kep}\,\ton{t-t_\mathrm{p}},
\eqf
it turns out that
\eqi
\eta = n_\mathrm{Kep}\,\ton{t_0 - t_\mathrm{p}}.
\eqf
Thus, since $n_\mathrm{Kep}$ stays constant because the semimajor axis is not secularly affected by the gravitomagnetic field, the rate of change of the mean anomaly at epoch is proportional to the opposite of the pace of variation of the time of passage at pericenter according to
\eqi
\dert\eta t = -n_\mathrm{Kep}\,\dert{t_\mathrm{p}} t.
\eqf
Should $\eta$ increase, the crossing of the pericenter would be anticipated with respect to the Keplerian case since $t_\mathrm{p}$ would decrease, and vice versa. In this case, the variation of $\eta$ would result in an orbit-by-orbit advance or delay of the passages at the pericenter, which does not occur in the present case because, in fact, $\mathrm{d}\eta/\mathrm{d}t=0$ for \rfr{ALT}. Instead, the 1pN gravitoelectric acceleration due to the mass monopole of the source affects $\eta$ with a negative rate corresponding to an increase of $t_\mathrm{p}$ \citep{2016MNRAS.460.2445I}.
}
\section{A gravitomagnetic clock effect for LAGEOS and LARES 2; is it possible to measure it?}\lb{sec:7}
Strengthened by the numerical confirmations obtained in the previous Sections of our recent analytical results for the pK orbital periods \citep{2016MNRAS.460.2445I}, we can trustworthily apply them to some specific situations of interest.

It turns out that a form of gravitomagnetic clock effect can be derived for the laser-ranged geodetic satellites LAGEOS (L) \citep{1985JGR....90.9217C} and LARES 2 (LR 2) \citep{2019JGeod..93.2437P} in view of their peculiar orbital configuration characterized by essentially identical circular orbits lying in orbital planes inclined to the Earth's equator  by
\begin{align}
I_\mathrm{L} \lb{IL} & \simeq 110^\circ,\acap
I_\mathrm{LR\,2} \lb{ILR2} &\simeq \pi - I_\mathrm{L}=70^\circ,
\end{align}
respectively. Indeed, although none of them move in the Earth's equatorial plane, their supplementary inclinations make them feasible, at least in principle, to detect the difference of the gravitomagnetic corrections to their sidereal periods meant as the time intervals between two consecutive crossings of a fixed reference direction in space.

By aligning $\boldsymbol{J}$ along the reference $z$ axis of a coordinate system whose fundamental plane is parallel to the Earth's equator, Equation\,(76) and Equation\,(78) of \citet{2016MNRAS.460.2445I}, valid to zero order in $e$, yield
\eqi
\delta T_\mathrm{gvm}^\mathrm{sid} = 4\,\pi\,\ton{-1 + 3\,\cos I}\,\rp{J}{M\,c^2}.\lb{tsid}
\eqf
By calculating \rfr{tsid} with \rfrs{IL}{ILR2} one obtains
\eqi
\Delta T_\mathrm{gvm}^\mathrm{sid}:=\delta T_\mathrm{gvm}^\mathrm{sid\,L} - \delta T_\mathrm{gvm}^\mathrm{sid\,LR\,2}  = 24\,\pi\,\cos I_\mathrm{L}\,\rp{J}{M\,c^2} = -2.8\times 10^{-7}\,\mathrm{s}.\lb{llrLT}
\eqf

Unfortunately, the mutual cancelation of the Keplerian orbital periods $P_\mathrm{Kep}$, which should be ideally identical, is far from being perfect. Indeed, for \citep{2023EPJC...83...87C}
\begin{align}
a_\mathrm{L} & = 12270.020705\,\mathrm{km}, \acap
a_\mathrm{LR\,2} & = 12266.1359395\,\mathrm{km},
\end{align}
it is
\eqi
\Delta P_\mathrm{Kep}:= P_\mathrm{Kep}^\mathrm{L} - P_\mathrm{Kep}^\mathrm{LR\,2} = \rp{2\,\pi}{\sqrt{\mu_\oplus}}\,\ton{\sqrt{a^3_\mathrm{L}} - \sqrt{a^3_\mathrm{LR\,2}}} = 6.4\,\mathrm{s}.
\eqf
What is worse, $\Delta P_\mathrm{Kep}$ cannot be even known with good enough accuracy to allow it to be safely subtracted from the measured difference of the satellites' orbital periods.
Indeed, since the currently claimed uncertainties in the spacecraft's semimajor axes are
\citep{2023EPJC...83...87C}
\begin{align}
\sigma_{a_\mathrm{L}} \lb{saL}& = 10^{-6}\,\mathrm{km} = 1\,\mathrm{mm}, \acap
\sigma_{a_\mathrm{LR\,2}} \lb{saLR2}& = 10^{-7}\,\mathrm{km} = 0.1\,\mathrm{mm},
\end{align}
the uncertainty in the difference of the Keplerian periods is as large as
\eqi
\sigma_{\Delta P_\mathrm{Kep}} = 1.6\times 10^{-6}\,\mathrm{s},
\eqf
which is about six times larger than the gravitomagnetic clock effect of \rfr{llrLT}. Incidentally, the Earth's gravitational parameter $\mu_\oplus$ would not be a limiting factor since its error \citep{iers10} $\sigma_{\mu_\oplus} = 8\times 10^5\,\mathrm{m^3\,s}^{-2}$ contributes to $\sigma_{\Delta P_\mathrm{Kep}}$ at the $\simeq 6\times 10^{-9}\,\mathrm{s}$ level.
Only an improvement of the already unrealistically small errors of \rfrs{saL}{saLR2} by three orders of magnitude would allow to bring down $\sigma_{\Delta P_\mathrm{Kep}}$ to the $\simeq 10^{-9}\,\mathrm{s}$ level.
\section{Summary and conclusions}\lb{sec:8}
A recently published analytical calculation  of the 1pN gravitomagnetic correction to the draconitic period of a test particle moving along a circular orbit arbitrarily oriented with respect to the equatorial plane of a massive, spinning primary can be specialized to the case in which the angular momentum $\boldsymbol{J}$ of the latter, on whose orientation in space no constraints are assumed as well, is (anti)parallel to the orbital angular momentum $\boldsymbol{h}$. In such a scenario, in which the orbital plane lies in the equatorial one of the source, the line of the nodes is left unaffected by the pN gravitomagnetic field of the primary, and can be naturally used as reference axis in the orbital plane from which the polar angle $\upvarphi$, represented in this case by the argument of latitude $u$, is reckoned. It turns out that the corrections $\delta T^\pm_\mathrm{gvm}$, to be added to the Keplerian periods $P_\mathrm{Kep}$ in order to have the time intervals between two consecutive crossings of the line of the nodes, are larger than the ones present so far in the literature, equal to $\pm 2\pi\,J/\ton{M\,c^2}$ depending on the prograde or retrograde directions, respectively, by a factor of 4 in both the senses of motion. Also the amount by which the angle $\overline{\upvarphi}$ marking the encounter of the two counter-rotating particles is reduced with respect to the Keplerian case is 4 times larger than the value quoted in the literature.

All such  features are fully confirmed by numerical integrations of the equations of motion of a fictitious test particle orbiting a putative massive, spinning object and experiencing, to the 1pN order, the gravitomagnetic acceleration ${\bds A}_\mathrm{LT}$ induced by $\boldsymbol{J}$ in addition to the Newtonian inverse-square one ${\bds A}_\mathrm{N}$. In particular, the time required by the cosine of the angle between the satellite's position vector $\boldsymbol{r}$ and the unit vector $\boldsymbol{\hat{l}}$ of the line of the nodes to take its initial value again, assumed equal to unity, turns out to be given by the Keplerian orbital period increased or decreased by $8\pi\,J/\ton{M\,c^2}$ for the prograde or retrograde senses of motion, respectively. Such a discrepancy with respect to the commonly known scenario may be due to the fact that the latter is based on the strict validity of the circularity condition throughout the integration over one orbital revolution, not allowing any instantaneous variations of the orbital parameters entering the analytical expression of the time derivative of the polar angle. Such a feature is, instead, accounted for in the aforementioned recent calculation which takes into account also the possible motion of the line of the nodes itself occurring, in general, for a non-equatorial orbits.
As a result, the revised gravitomagnetic clock effect for identical circular and equatorial orbits traveled in opposite directions turns out to be $16\pi\,J/\ton{M\,c^2}$. \textcolor{black}{Such a result has potential relevance to astrophysical observations of the environment of compact objects around which the gravitomagnetic field plays a major roles.}

The numerical experiments performed in the present work confirmed also another analytically inferred distinctive feature of the temporal structure around a rotating body: there are no gravitomagnetic corrections to the anomalistic periods for both the senses of motion. \textcolor{black}{It is so because the time of passage at pericenter is left unaffected by the gravitomagnetic acceleration, contrary to the 1pN gravitoelectric one.}

Although they do not share the same orbital plane nor do any of them move in the Earth's equatorial plane, the artificial satellites LAGEOS and LARES 2 may be used, at least in principle, to detect a form of clock effect involving the difference of their sidereal orbital periods which is, ideally, entirely gravitomagnetic, amounting to $\simeq 3\times 10^{-7}\,\mathrm{s}$. Actually, despite the close values of their semimajor axes, their Keplerian orbital periods do not cancel each other, being their difference uncertain at a $\simeq 10^{-6}\,\mathrm{s}$ level due to the current claimed errors in the aforementioned orbital elements.
\begin{appendices}
\section{Notations and definitions}\lb{appenA}
Here, some basic notations and definitions used throughout the text are presented \citep{Sof89,1991ercm.book.....B,2003ASSL..293.....B,2011rcms.book.....K,2014grav.book.....P,SoffelHan19}.
\begin{itemize}
\item[] $c:$ speed of light in vacuum
\item[] $G:$ Newtonian constant of gravitation
%\item[] $g_{\sigma\nu}:$ spacetime metric tensor
%\item[] $\phi,~w:$ gravitoelectric potential
%\item[] $U:$ Newtonian gravitational potential
%\item[] $\mathbf{w}:$ gravitomagnetic potential
%\item[] $T^{\sigma\nu}:$ energy-momentum tensor of the source
%\item[] $\epsilon:$ mean obliquity
%\item[] $\upmu_0:$ magnetic permeability of vacuum
\item[] $M:$ mass of the central body
%\item[] $M_\odot:$ mass of the Sun
\item[] $\mu:= GM:$ gravitational parameter of the central body
\item[] $\boldsymbol{J}:$ angular momentum of the central body
\item[] $J:$ magnitude of the angular momentum of central body
%\item[] $\chi_g:$ dimensionless spin parameter of a Kerr black hole: it is $\left|\chi_g\right|\leq 1$.
\item[] $\alpha_J:$ right ascension (RA) of the north pole of rotation of the central body
\item[] $\delta_J:$ declination (DEC) of the north pole of rotation of the central body
\item[] ${\boldsymbol{\hat{J}}}=\grf{\cos\alpha_J\,\cos\delta_J,\,\sin\alpha_J\,\cos\delta_J,\,\sin\delta_J}:$ spin axis of the central body
%\item[] ${\hat{k}}_x = \cos\alpha_J\,\cos\delta_J:$ $x$ component of the  spin axis of the central body
%\item[] ${\hat{k}}_y = \sin\alpha_J\,\cos\delta_J:$ $y$ component of the  spin axis of the central body
%\item[] ${\hat{k}}_z = \sin\delta_J:$ $z$ component of the  spin axis of the central body
%\item[] $R_\textrm{e}:$ equatorial radius of the central body
%\item[] $R_\textrm{p}:$ polar radius of the central body
%\item[] $\varepsilon:=\sqrt{1 - \ton{\rp{R_\textrm{p}}{R_\textrm{e}}}^2}:$ ellipticity of the central body
%\item[] $J_2:$ dimensionless zonal harmonic coefficient of degree $\ell=2$ of the non spherically symmetric gravitational potential of the central body
%\item[] $C_{2,1},~S_{2,1},~C_{2,2},~S_{2,2}:$ tesseral and sectorial multipole mass moments of degree $\ell = 2$ of Jupiter
%\item[] $Q_2:$ dimensional mass quadrupole moment of the non spherically symmetric gravitational potential of the central body
%\item[] ${\bds B}^\mathrm{oct}:$ gravitomagnetic spin-octupole field in the empty space surrounding the rotating central body
%\item[] $\phi^\mathrm{oct}:$ gravitomagnetic spin-octupole potential in the empty space surrounding the rotating central body
\item[] ${\bds A}:$ perturbing acceleration experienced by the test particle
\item[] ${\bds A}_\mathrm{N}:$ Newtonian inverse-square acceleration
%\item[] $\alpha_\textrm{X}:$ right ascension (RA) of the 3rd body's spin axis
%\item[] $\delta_\textrm{X}:$ declination (DEC) of the 3rd body's spin axis
%\item[] $\kx^\textrm{eq}=\cos\delta_\textrm{X}\cos\alpha_\textrm{X}:$ component of the 3rd body's spin axis with respect to  the reference $x$ axis of an equatorial %coordinate system
%\item[] $\ky^\textrm{eq}=\cos\delta_\textrm{X}\sin\alpha_\textrm{X}:$ component of the 3rd body's spin axis with respect to  the reference $y$ axis of an equatorial %coordinate system
%\item[] $\kz^\textrm{eq}=\sin\delta_\textrm{X}:$ component of the 3rd body's spin axis with respect to  the reference $z$ axis of an equatorial coordinate system
\item[] $\upvarphi: $ azimuthal angle reckoning the instantaneous position of the test particle in its orbital plane
\item[] $\upomega:= \mathrm{d}\upvarphi/\mathrm{d}t:$ azimuthal angular speed of the test particle in its orbital plane
\item[] $T_\mathrm{gvm}:$ orbital period in presence of the LT acceleration
\item[] $a:$  semimajor axis of the test particle
\item[] $\nk := \sqrt{\mu/a^3}:$  Keplerian mean motion of the test particle
\item[] $\Pb:= 2\uppi/\nk:$ orbital period of the test particle
\item[] $e:$  eccentricity of the test particle
\item[] $p:= a\ton{1-e^2}:$ semilatus rectum of the orbit of the test particle
\item[] $I:$  inclination of the orbital plane of the test particle to the reference plane $\grf{x,\,y}$
\item[] $\Omega:$  longitude of the ascending node  of the test particle
\item[] $\ascnode:$ ascending node
\item[] $\descnode:$ descending node
\item[] $\omega:$  argument of pericenter  of the test particle
\item[] $q: = e\,\cos\omega: $ non-singular orbital element $q$
\item[] $k: = e\,\sin\omega: $ non-singular orbital element $k$
%\item[] $\eta:$ mean anomaly at epoch
\item[] $f\ton{t}:$ true anomaly of the test particle
%\item[] $f_0:$ true anomaly of the test particle at some arbitrary moment of time $t_0$ assumed as initial instant
\item[] $u:= \omega + f$ argument of latitude of the test particle
\item[] \textcolor{black}{$\eta:$ mean anomaly at epoch}
\item[] \textcolor{black}{$t_0:$ initial instant of time}
\item[] \textcolor{black}{$t_\mathrm{p}:$ time of passage at pericenter}
\item[] \textcolor{black}{$\mathcal{M}\ton{t}$: mean anomaly}
\item[] $\boldsymbol{r}:$ position vector of the test particle with respect to the central body
\item[] $r:$ distance of the test particle from the central body
\item[] $r_0:$ radius of a circular orbit
\item[] ${\boldsymbol{\hat{r}}}:= {\boldsymbol{r}}/r = \grf{\cos\Omega\,\cos u - \cos I\,\sin\Omega\,\sin u,\, \sin\Omega\,\cos u + \cos I\,\cos\Omega\,\sin u,\,\sin I\,\sin u}:$ radial unit vector
%\item[] $\xi:= \boldsymbol{\hat{J}}\boldsymbol{\cdot}\boldsymbol{\hat{r}}:$ cosine of the angle between the central body's spin axis and the position vector of the test particle
%\item[] $\delta:$ declination (DEC) of the test particle
\item[] $\boldsymbol{v}:$ velocity vector of the test particle
%\item[] $v_r:=\boldsymbol{v}\boldsymbol{\cdot}\boldsymbol{\hat{r}}:$ radial velocity of the test particle
%\item[] $\lambda := \mathbf{\hat{k}}\mathbf{\cdot}\mathbf{v}:$ projection of the velocity of the test particle onto the direction of the spin axis %of the central body
%\item[] $\mathcal{P}_{\ell}\ton{\cdots}:$ Legendre polynomial of degree $\ell$
\item[] $\boldsymbol{h} = \boldsymbol{r}\boldsymbol{\times}\boldsymbol{v}:$ orbital angular momentum per unit mass
\item[] $\boldsymbol{\hat{h}}:=\grf{\sI\sO,~-\sI\cO,~ \cI}:$ unit vector of the orbital angular momentum such that $\boldsymbol{\hat{l}}\boldsymbol{\times}\boldsymbol{\hat{m}}=\boldsymbol{\hat{h}}$
\item[] \textbf{\textit{A}}$=\boldsymbol{v}\boldsymbol{\times}\boldsymbol{h} - \mu\,\boldsymbol{\hat{r}}$: Laplace-Runge-Lenz vector per unit mass
\item[] $\boldsymbol{\hat{l}}:=\grf{\cO,~\sO,~0}:$ unit vector directed along the line of the nodes toward the ascending node
\item[] $\boldsymbol{\hat{m}}:=\grf{-\cI\sO,~\cI\cO,~\sI}:$ unit vector directed transversely to the line of the nodes in the orbital plane
\item[] $\boldsymbol{\hat{s}}:=\boldsymbol{\hat{h}}\boldsymbol{\times}\boldsymbol{\hat{r}}=\grf{-\sin u\,\cos\Omega - \cos I\,\sin\Omega\,\cos u,\,-\sin\Omega\,\sin u + \cos I\,\cos\Omega\,\cos u,\,\sin I\,\cos u}:$ transverse unit vector
\item[] $\mathrm{A}_R:= \bds A\boldsymbol{\cdot}\boldsymbol{\hat{r}}:$ radial component of the perturbing acceleration $\bds A$
\item[] $\mathrm{A}_T:= \bds A\boldsymbol{\cdot}\boldsymbol{\hat{s}}:$ transverse component of the perturbing acceleration $\bds A$
\item[] $\mathrm{A}_N:= \bds A\boldsymbol{\cdot}\boldsymbol{\hat{h}}:$ normal component of the perturbing acceleration $\bds A$
%\item[] $\boldsymbol{\hat{P}}:= \boldsymbol{\hat{l}}\cos\omega + \boldsymbol{\hat{m}}\sin\omega:$ unit vector in the orbital plane directed along the line of apsides towards %the pericenter
%\item[] $\boldsymbol{\hat{Q}}:= -\boldsymbol{\hat{l}}\sin\omega + \boldsymbol{\hat{m}}\cos\omega:$ unit vector in the orbital plane directed transversely to the line of %apsides
\end{itemize}
\end{appendices}
\section*{Data availability}
No new data were generated or analysed in support of this research.
\section*{Conflict of interest statement}
I declare no conflicts of interest.
\bibliography{Uranusbib}{}
\end{document}